\documentclass[prd,aps,twocolumn,preprintnumbers,showpacs,nofootinbib,superscriptaddress,notitlepage]{revtex4-1}

\usepackage{hyperref}
\usepackage{mathrsfs}
\usepackage{amsmath}
\usepackage{slashed}
\usepackage{graphicx}
\usepackage{xcolor}
\usepackage{xspace}

\usepackage{amssymb}
\usepackage[utf8]{inputenc}
\usepackage[normalem]{ulem}

\newcommand{\beq}{\begin{eqnarray}}
\newcommand{\eeq}{\end{eqnarray}}

\newcommand{\nn}{\nonumber}

\newcommand{\spcorr}{spatial correlator\xspace}

\DeclareRobustCommand{\app}[1]{App.~\ref{sec:#1}}

\DeclareRobustCommand{\eq}[1]{Eq.~(\ref{eq:#1})}
\DeclareRobustCommand{\eqs}[2]{Eqs.~(\ref{eq:#1}) and (\ref{eq:#2})}
\DeclareRobustCommand{\sec}[1]{Sec.~\ref{sec:#1}}

\DeclareRobustCommand{\figs}[2]{Figs.~\ref{fig:#1} and \ref{fig:#2}}
\DeclareRobustCommand{\app}[1]{App.~\ref{sec:#1}}

\begin{document}
	
\preprint{MIT-CTP 4960}
	
\title{Factorization Theorem Relating Euclidean and Light-Cone Parton Distributions}
	
\author{Taku Izubuchi}
\affiliation{Physics Department, Brookhaven National Laboratory, Upton, NY, 11973, USA}
\affiliation{RIKEN-BNL Research Center, Brookhaven National Laboratory, Upton, NY 11973, USA}

\author{Xiangdong Ji}
\affiliation{Tsung-Dao Lee Institute, and College of Physics and Astronomy,
Shanghai Jiao Tong University, Shanghai, 200240, P. R. China}
\affiliation{Maryland Center for Fundamental Physics, Department of Physics,
	University of Maryland, College Park, Maryland 20742, USA}

\author{Luchang Jin}
\affiliation{Physics Department, Brookhaven National Laboratory, Upton, NY, 11973, USA}
\affiliation{RIKEN-BNL Research Center, Brookhaven National Laboratory,
	Upton, NY 11973, USA}
\affiliation{Physics Department, University of Connecticut,
Storrs, Connecticut 06269-3046, USA}

\author{Iain W. Stewart}
\affiliation{Center for Theoretical Physics, Massachusetts Institute of Technology, Cambridge, MA 02139, USA}

\author{Yong Zhao\vspace{0.2cm}}
\affiliation{Center for Theoretical Physics, Massachusetts Institute of Technology, Cambridge, MA 02139, USA}

\begin{abstract}
	
\vspace{0.05cm}
In a large-momentum nucleon state, the matrix element of a gauge-invariant Euclidean Wilson line operator accessible from lattice QCD can be related to the standard light-cone parton distribution function through the large-momentum effective theory (LaMET) expansion. This relation is given by a factorization formula with a non-trivial matching coefficient.  Using the operator product expansion we derive the large-momentum factorization of the quasi-parton distribution function in LaMET, and show that the more recently discussed pseudo-parton distribution approach also obeys an equivalent factorization formula.  Explicit results for the coefficients are obtained and compared at one-loop.  We also prove that the matching coefficients in the $\overline{\rm MS}$ scheme depend on the large partonic momentum rather than the nucleon momentum.	
\end{abstract}

\maketitle

\section{Introduction}

Parton distribution functions (PDFs) are key quantities for gaining an understanding of hadron structure and for making predictions for the cross sections in high-energy scattering experiments. In QCD factorization theorems for hard scattering processes~\cite{Collins:1989gx}, the relevant PDFs are defined in terms of the nucleon matrix elements of light-cone correlation operators. For example, in dimensional regularization with $d=4-2\epsilon$, the bare unpolarized quark PDF is
\begin{align} \label{eq:lcpdf}
	q(x,\epsilon) \equiv\! \int\! {d\xi^-\over 4\pi} e^{-ixP^+\xi^-}\! \langle P | \bar{\psi}  (\xi^-) \gamma^+ U (\xi^-, 0) \psi (0) | P \rangle ,
\end{align}
where $x$ is the momentum fraction, the nucleon momentum $P^\mu=(P^0,0,0,P^z)$, $\xi^{\pm}=(t\pm z)/\sqrt{2}$ are the light-cone coordinates, and the Wilson line is
\begin{align}
	U(\xi^-,0) = P\exp\biggl(-ig\int_0^{\xi^-} d\eta^- A^+(\eta^-)\biggr)\ .
\end{align}
Most often the bare PDF is renormalized in the $\overline{\ensuremath{\operatorname{MS}}}$ scheme to obtain $q(x,\mu)$, and this renormalized PDF  is used to make predictions for experiment. The relation is
\beq
	q(x,\epsilon) = \int_x^1 {dy\over y}\:  Z^{\overline{\ensuremath{\operatorname{MS}}}}\Bigl({x\over y},\epsilon,\mu\Bigr) q(y,\mu)\ ,
\eeq
where $\mu$ is the renormalization scale, and we have suppressed the flavor indices in the renormalization constant $Z^{\overline{\ensuremath{\operatorname{MS}}}}$ and PDFs. In light-cone quantization with $A^+=0$, the $\overline{\ensuremath{\operatorname{MS}}}$ definition has an interpretation as a parton number density.

So far our main knowledge of the PDF is obtained from global fits to deep inelastic scattering and jet data, see for example~\cite{Ball:2014uwa,Dulat:2015mca,Martin:2009iq,Alekhin:2017kpj,Buckley:2014ana}. On the other hand, calculating the PDF from first principles with QCD has been an attractive subject, which can for example provide access to spin and momentum distributions that are hard to determine experimentally. Several different approaches to this have been considered using the lattice theory which is a nonperturbative method to solve QCD. Since the lattice theory is defined in a discretized Euclidean space with imaginary time, it is very difficult to calculate Minkowskian quantities with real-time dependence such as the PDF. The first and most well explored option is calculating the moments of the PDF~\cite{Martinelli:1987zd,Martinelli:1988xs,Detmold:2001dv,Detmold:2002nf,Dolgov:2002zm} that are matrix elements of local gauge-invariant operators. However, since the lattice regularization breaks $O(4)$ rotational symmetry, the consequent mixing between operators of different dimensions makes it difficult to compute higher moments, which in practice has limited the amount of information that can be extracted from this approach. A method to improve this situation by restoring the rotational symmetry has been proposed in Ref.~\cite{Davoudi:2012ya}.
Other proposals include extracting the PDF from the hadronic tensor~\cite{Liu:1993cv,Liu:1998um,Liu:1999ak,Liu:2016djw} and the forward Compton amplitude~\cite{Chambers:2017dov}, possibly with flavor changing currents~\cite{Detmold:2005gg}, and the more general ``lattice cross sections"~\cite{Ma:2014jla,Ma:2017pxb}. Systematic lattice analyses of these approaches are  under investigation, but challenges remain.

In Ref {\cite{Ji:2013dva}} Ji proposed that the $x$-dependence of the PDF can be extracted from a Euclidean distribution on the lattice, which can be understood in the language of the large momentum effective theory (LaMET)~\cite{Ji:2014gla}. This Euclidean distribution is referred to as the quasi-PDF, whose bare matrix element is defined using a spatial correlation of quarks along the $z$ direction,
\beq \label{eq:qpdf}
	\tilde{q}(x,P^z,\epsilon) \equiv \int_{-\infty}^\infty {dz\over 4\pi} e^{ixP^zz}\langle P | \bar{\psi}  (z) \Gamma U (z, 0) \psi (0) | P \rangle \,,\nn\\
\eeq
where $\Gamma=\gamma^z$, $z^\mu=ze^\mu$, $e^\mu=(0,0,0,1)$, and the Wilson line is
\beq
	U(z,0) = P\exp\left(-ig\int_0^z dz' A^z(z')\right)\ .
\eeq
For finite momentum $P^z$, $\tilde{q}(x,P^z,\epsilon)$ has support in $-\infty < x < \infty$.
According to Ref.~\cite{Hatta:2013gta}, there is a universality class of operators that can be considered. For example, for the quasi-PDF, one could also replace $\Gamma=\gamma^z$ by $\Gamma=\gamma^0$ in Eq.~(\ref{eq:qpdf}) as both definitions reduce to the PDF under an infinite Lorentz boost along the $z$ direction. Unlike the PDF in Eq.~(\ref{eq:lcpdf}) that is invariant under the Lorentz boost, the quasi-PDF depends dynamically on it through the nucleon momentum $P^z$. When the nucleon momentum $P^z$ is much larger that the nucleon mass $M$ and $\Lambda_{\rm QCD}$, which is an attainable window on the lattice, the quasi-PDF can be factorized
into a matching coefficient and the PDF~\cite{Ji:2013dva,Ji:2014gla}. The factorization formula is
\begin{align} \label{eq:momfact}
	\tilde{q}(x,P^z,\mu_R)
	=& \int_{-1}^1 {dy\over |y|}\: C\Bigl({x\over y},{\mu_R\over \mu},{\mu\over p^z}\Bigr) \, q(y,\mu)
	\nn\\
	& + \mathcal{O}\biggl({M^2\over P_z^2},{\Lambda_{\ensuremath{\operatorname{QCD}}}^2\over P_z^2}\biggr)\ ,
\end{align}
where the renormalized quasi-PDF $\tilde{q}(x,P^z,\mu_R)$ is defined in a particular scheme at renormalization scale $\mu_R$, and $\mathcal{O}(M^2/P_z^2, \Lambda_{\rm QCD}^2/P_z^2)$ are power corrections suppressed by the nucleon momentum. In general the result for $C$ will depend on the choice of $\Gamma=\gamma^z$ or $\gamma^0$ and
renormalization schemes. For $\Gamma=\gamma^z$ the matching coefficient $C$ has been computed for the iso-vector quark quasi-PDF at one-loop level, first with a transverse momentum cutoff in~\cite{Xiong:2013bka}, confirmed in~\cite{Ma:2014jla,Alexandrou:2015rja}, and also recently determined in the regularization-invariant momentum subtraction (RI/MOM) scheme~\cite{Stewart:2017tvs}. Matching for the gluon quasi-PDF is calculated in~\cite{Wang:2017qyg,Wang:2017eel}. The matching coefficient $C$ is independent of the choice of states used for $\tilde q$ and $q$.\footnote{The scheme definition itself may separately involve a choice of state, such as in the RI/MOM scheme, but the result is still an operator renormalization that can be used for different choices of hadronic states for $\tilde q$ and $q$. The transverse cutoff and $\overline{\rm MS}$ schemes do not require even this choice.} Since matching calculations are carried out with quark states of momentum $p^z$, it can be tricky to know what the right choice to make is for $C$, and in some of the literature the choice of $p^z=P^z$ has been suggested when utilizing $C$ for the hadronic nucleon state. This is for example the case in the original quasi-PDF papers~\cite{Ji:2013dva,Ji:2014gla,Xiong:2013bka} and in the pioneering lattice calculations of the PDF from the quasi-PDF in~\cite{Lin:2014zya,Alexandrou:2015rja,Chen:2016utp,Alexandrou:2016jqi,Zhang:2017bzy,Alexandrou:2017huk,Chen:2017mzz,Lin:2017ani,Chen:2017gck}, which was summarized in Ref.~\cite{Lin:2017snn}. In the quasi-generalized parton distribution analysis in~\cite{Ji:2015qla} it was observed that one should take $p^z=|y|P^z$.  Through our rigorous analysis of \eq{momfact} we show that the correct result for this equation is indeed $p^z=|y|P^z$.

Recently, a different procedure~\cite{Radyushkin:2017cyf} to extract PDFs from the same lattice QCD matrix element as in~\cite{Ji:2013dva}
has been proposed based on the Lorentz invariant variables of the \spcorr (or pseudo-PDF), in place of the quasi-PDF. In this approach, one starts from the \spcorr $\tilde{Q}_{\gamma^\mu}$ defined for $\mu=0$ or $\mu=z$ by
\begin{align}
	{1\over2}\langle P | \tilde{O}_{\gamma^\mu}(z,\epsilon) | P\rangle
	= & P^\mu \tilde{Q}_{\gamma^\mu}(\zeta=P^zz,z^2,\epsilon) \,,	
\end{align}
which depends on the two Lorentz invariants $z^2$ and $\zeta=-z\cdot P = P^z z$; the latter is also called the Ioffe time.
For an arbitrary Dirac matrix $\Gamma$ the operator $\tilde{O}_\Gamma$ is defined as
\begin{align} \label{eq:OGamma}
	\tilde{O}_{\Gamma}(z,\epsilon)  = \bar{\psi}  (z) \Gamma U (z, 0) \psi (0) \,.
\end{align}
This is the same spatial correlator (calculable on lattice) used to define the quasi-PDF in \eq{qpdf}, where $P^z$ is fixed and one Fourier transforms with respect to $z$. If instead $z^2$ is fixed, and we Fourier transform from the Ioffe time $\zeta$---which is in principle integrating over $P^z$---to the momentum fraction $x$, then one obtains the pseudo-PDF~\cite{Radyushkin:2017cyf},
\begin{align} \label{eq:pseudo}
\mathcal{P} \left( x, z^2,\epsilon\right)
   = \int_{-\infty}^\infty \! \frac{d \zeta}{2\pi}\:
    e^{ix \zeta}\: \tilde{Q}_{\gamma^0}\bigl( \zeta, z^2,\epsilon \bigr)
   \,.
\end{align}
For arbitrary finite $z$, the pseudo-PDF only has support in $-1\le x \le 1$~\cite{Radyushkin:1983wh,Radyushkin:2016hsy}, but has no parton model interpretation.
(Again the pseudo-PDF can equally well be considered for $\Gamma=\gamma^z$.) The \spcorr or pseudo-PDF approach has been explored on the lattice~\cite{Orginos:2017kos,Karpie:2017bzm}, where the short distance behavior was explored.
The PDF corresponds to the situation when $z^\mu$ is light-like, in which case the space-time correlator is referred to as the Ioffe-time distribution~\cite{Ioffe:1969kf},
\begin{align}
 Q(\zeta,\epsilon) = \tilde Q_{\gamma^+}(\zeta=-P^+\xi^-,z^2=0,\epsilon)
  \,.
\end{align}
When Fourier transformed this correlation gives the PDF
\begin{align} \label{eq:Qdefn}
q(y,\epsilon) =\int_{-\infty}^{\infty} {d\zeta\over 2\pi} \ e^{iy\zeta}\: Q(\zeta,\epsilon)\,.
\end{align}
In short the quasi-PDF and pseudo-PDF are different representations of the Euclidean \spcorr, as summarized in Table.~\ref{tab:I}.
\begin{table}
\begin{center}
	\begin{tabular}{ | c | c | c |}
		\hline
		Distribution & Fourier transform & Arguments\\[-2pt]
		& from \spcorr & \\ \hline
		Spatial correlation  &  & $\zeta = zP^z, z^2$\\ \hline
		Quasi-PDF & $z\to xP^z$ & $x, P^z$\\ \hline
		Pseudo-PDF & $\zeta \to x$ & $x, z^2$\\ \hline
	\end{tabular}
\caption{Summary of the relationship between different Euclidean distributions.}
\label{tab:I}
\end{center}
\end{table}

It was pointed out in Ref.~\cite{Ji:2017rah} that to obtain enough information to extract the PDF for the \spcorr with small $z^2$, one has to do lattice calculations with large momenta $P^z$, which is the same requirement as for the quasi-PDF. Ref.~\cite{Ji:2017rah} also proposed that the renormalized pseudo-PDF satisfies the following small $z^2$ factorization,
\begin{align} \label{eq:pseudofact}
	\mathcal{P} \left( x, z^2\mu_R^2\right)
  &= \int {dy\over |y|}\: \mathcal{C}\left({x\over y}, {\mu_R^2\over \mu^2},z^2\mu_R^2\right) q(y,\mu)
   \nn\\
  &\ + {\cal O}(z^2\Lambda_{\rm QCD}^2, z^2 M^2) \,,
\end{align}
which they verified at order $O(\alpha_s)$ for the unpolarized iso-vector case with $\Gamma=\gamma^0$.  (Again the coefficient ${\cal C}$ will depend on the choice of $\Gamma=\gamma^0$ or $\gamma^z$.)

In Ref.\cite{Ma:2014jla} a diagrammatic derivation of the factorization formula in \eq{momfact} for the quasi-PDF was given.  Here we derive this factorization formula for the quasi-PDF in an alternate manner, and also show that \spcorr and pseudo-PDF are different representations of the same
fundamental factorization. Our approach is based on the operator product expansion (OPE) for spacelike separated local operators~\cite{Wilson:1969zs}. For such operators the OPE has been proven for scalar field theory to all orders in perturbation theory~\cite{Brandt:1967rb,Wilson:1972ee,Zimmermann:1972tv}, and is widely assumed to hold for any renormalizable quantum field theory including QCD. By introducing auxiliary fields in place of the Wilson line~\cite{Dorn:1986dt}, the correlator in \eq{OGamma} is known to be equivalent to a product of local renormalizable operators of this type. Through our derivation we find the explicit form of the large $P^z$ and small $z^2$ factorization formulas in Eq.(\ref{eq:momfact}) and Eq.~(\ref{eq:pseudofact}) respectively, as well as the relationship between the matching coefficients $C$ and $\mathcal{C}$. Since the requirement for large $P^z$ and small $z^2$ is the same for both the quasi-PDF and pseudo-PDF approaches, there is in principle no fundamental difference in applying either one
to lattice calculations of the proton matrix element of $\tilde{O}_{\Gamma}(z)$. It is interesting to compare
both approaches utilizing the same lattice data, although they shall not yield different result in principle.

The rest of this paper is organized as follows: In Sec.~\ref{sec:ope}, we use an OPE of $\tilde{O}_\Gamma(z)$ to derive the large $P^z$ factorization of LaMET in Eq.~(\ref{eq:momfact}) and small $z^2$ factorization of the pseudo-PDF in Eq.~(\ref{eq:pseudofact}). We prove that one must take $p^z=|y|P^z$ in \eq{momfact}, so the corresponding argument in $C$ is $\mu/(|y|P^z)$.
(This OPE approach was used recently in Ref.~\cite{Ma:2017pxb} to prove the factorization theorem for the ``lattice cross sections", and the OPE proof carried out here was done independently and first presented in Ref.~\cite{Zhao:2017int}.)
In Sec.~\ref{sec:equiv}, we derive the \spcorr, pseudo-PDF, and quasi-PDF distributions and matching coefficients at one-loop in $\overline{\rm MS}$ and analyze the Fourier-transform relation between the quasi-PDF and pseudo-PDF. Unlike earlier results for the quasi-PDF in $\overline{\rm MS}$, we also use dimensional regularization with minimal subtraction to renormalize divergences at $x=\pm\infty$. In Sec.~\ref{sec:ren}, we discuss how renormalization schemes other than $\overline{\rm MS}$ are easily incorporated into the factorization formulas.
In Sec.~\ref{sec:num} we carry out a numerical analysis of the matching coefficients, by computing the convolution in Eq.~(\ref{eq:momfact}) numerically using the PDF determined by global fits~\cite{Martin:2009iq}. We show that the difference between using $p^z=P^z$ and $p^z=|y|P^z$ in \eq{momfact} is an important effect, and that our $\overline{\rm MS}$ matching coefficients are insensitive to cutoffs in the convolution integral. In Sec.~\ref{sec:lattice}, we discuss the implications of our OPE analysis for the lattice calculation of the PDF in both the quasi-PDF and pseudo-PDF approaches. Finally, we conclude in Sec.~\ref{sec:summary}.

\section{Factorization from the OPE}
\label{sec:ope}

In this section we make use of the operator product expansion to derive the matching relation for the quasi-PDF, as well as the equivalent matching relations for the \spcorr and pseudo-PDF. For simplicity these three equivalent cases are presented in separate subsections.
 
\subsection{OPE and Factorization for the Spatial Correlator}

The OPE is a technique to expand nonlocal operators with separation $z^\mu$ in terms of local ones in the Euclidean limit of $z^2\to 0$.
It can be applied to both bare regulated operators as well as renormalized operators, and our focus will be on the latter. For the gauge-invariant Wilson operator $\tilde{O}_\Gamma(z)$, it was proven that it can be multiplicatively renormalized in  coordinate space as~\cite{Ji:2017oey,Ishikawa:2017faj}
\beq \label{eq:ren}
\tilde{O}_\Gamma(z,\mu) = Z_{\psi,z}\, e^{\delta m |z|} \tilde{O}_\Gamma(z,\epsilon)\,,
\eeq
where $\delta m$ captures the power divergence from the Wilson line self-energy, $Z_{\psi,z}$ only depends on the end points $z,0$ and renormalizes the logarithmic divergences. This multiplicative renormalization was also discussed earlier in Refs.~\cite{Ishikawa:2016znu,Chen:2016fxx,Constantinou:2017sej}.  For simplicity, in this section we take $\Gamma=\gamma^z$ for \eq{ren}.

In the $\overline{\rm MS}$ scheme, the power divergence vanishes, and using the OPE
the renormalized $\tilde{O}_\Gamma(z,\mu)$ can be expanded 
in terms of local gauge-invariant operators as $z^2\to 0$ giving
\begin{align}\label{eq:ope-tq}
\tilde{O}_{\gamma^z}(z,\mu) = & \sum_{n=0}^\infty \left[C_n ({\mu}^2 z^2)\frac{(-iz)^n}{n!} e_{\mu_1}\cdots 				e_{\mu_n}O_1^{\mu_0\mu_1\cdots\mu_n}(\mu)\right. \nn\\
&+C'_n ({\mu}^2 z^2)\frac{(-iz)^n}{n!} e_{\mu_1}\cdots e_{\mu_n}O_2^{\mu_0\mu_1\cdots\mu_n}(\mu)\nn\\
& +  \text{higher-twist operators}\Big]\,,
\end{align}
where $\mu_0=z$, $C_n=1+O(\alpha_s)$ and $C'_n=O(\alpha_s)$ are Wilson coefficients, and $O_1^{\mu_0\mu_1\cdots\mu_n}(\mu)$ and $O_2^{\mu_0\mu_1\cdots\mu_n}(\mu)$ are the only allowed renormalized traceless symmetric twist-2 quark and gluon operators at leading power in the OPE,
\begin{align}\label{eq:twist-2}
O_1^{{\mu}_0 {\mu}_1 \ldots {\mu}_n}(\mu) = & Z_{n+1}^{qq} \bigl(
 \bar{\psi} \gamma^{({\mu}_0 } iD^{{\mu}_1} \cdots iD^{ {\mu}_n)} \psi -
\text{trace} \bigr) \,,
  \\
O_2^{{\mu}_0 {\mu}_1 \ldots {\mu}_n}(\mu) = & Z_{n+1}^{qg}\bigl(
 F^{(\mu_0\rho} iD^{{\mu}_1} \cdots iD^{ {\mu}_{n-1}} F_\rho^{~\mu_n)} - \text{trace} \bigr)
  .\nn
\end{align}
Here $Z_{n+1}^{ij}=Z_{n+1}^{ij}(\mu,\epsilon)$ are multiplicative $\overline{\rm MS}$ renormalization factors and $(\mu_0 \cdots \mu_n)$ stands for the symmetrization of these Lorentz indices.

The above OPE is valid for the operator itself, where we implicitly  constrain ourselves to the subspace of matrix elements for which the twist expansion is appropriate. In the iso-vector case, the mixing with the gluon operators is absent, which we will stick to for the rest of the paper. When $O_1^{\mu_0\mu_1\cdots\mu_n}$ is evaluated in the nucleon state,
\beq \label{eq:def-moments}
\langle P | O_1^{{\mu}_0 {\mu}_1 \cdots {\mu}_n} | P\rangle =  2a_{n + 1}(\mu)\left(P^{{\mu}_0} P^{{\mu}_1} \ldots P^{{\mu}_n} - \text{trace}\right),  \nn\\
\eeq
where $a_{n+1}(\mu)$ is the $(n+1)$-th moment of the PDF,
\beq \label{eq:moment}
a_{n + 1} \left(\mu\right)=\int_{-1}^1 dx\,x^n q \left(x,\mu\right)\,,
\eeq
and the explicit expression of the trace term have been derived in Ref.~\cite{Nachtmann:1973mr,Georgi:1976ve,Chen:2016utp}. The inverse relation to \eq{moment} is that $q(x,\mu)$ has an expansion with terms proportional to the n'th derivative of the $\delta$-function, as in $\delta^{(n)}(x)\, a_n(\mu)$,  without any nontrivial short distance Wilson coefficient.

As pointed out in Ref.~\cite{Ji:2017rah}, to obtain enough information for the spatial correlator at $|\zeta|= |P^z z|\sim 1$ at small $z^2$, we have to choose $P^z$ large compared to the scale $\Lambda_{\rm QCD}$. When $P_z^2\gg\{\Lambda_{\rm QCD}^2, M^2\}$, the trace terms in Eq.(\ref{eq:def-moments}) are suppressed by powers of $M^2/P_z^2$, while the contributions from higher-twist operators in Eq.~(\ref{eq:ope-tq}) are suppressed by powers of $\Lambda_{\rm QCD}^2/P_z^2$ or $z^2\Lambda_{\rm QCD}^2$. Therefore, the twist-2 contribution is the leading approximation of the nucleon matrix element $\langle P|\tilde{O}_{\gamma^z}(z)|P\rangle$ at large momentum. From now on we will drop all the power corrections from our discussion.

The Wilson coefficients $C_n ({\mu}^2 z^2)$ in the OPE of $\tilde{O}_{\gamma^z}(z)$ can be computed in perturbation theory for $\mu\sim 1/|z|\gg \Lambda_{\rm QCD}$. In
the $\overline{\ensuremath{\operatorname{MS}}}$ scheme, the $C_n$ are log-singular near $z^2 =
0$, and so is $\langle P|\tilde{O}_{\gamma^z}(z,\mu)|P\rangle$. For this reason the $x$-moments of the quasi-PDF $\tilde q(x,P^z,\mu)$ are proportional to $C_n|_{z=0}$ which is
divergent, and the quasi-PDF will not simply become the PDF in the infinite $P^z$ limit. Instead, we need a factorization formula which matches the quasi-PDF to the PDF. In contrast, for the pseudo-PDF the moments do exist since we hold $\mu^2 z^2$ fixed when taking the $x$-moment. However, we still need a factorization formula to match the pseudo-PDF to the PDF.  We will comment further about this below.

Based on Eqs.~(\ref{eq:ope-tq}-\ref{eq:moment}), we can write down the leading-twist approximation to the spatial correlator as
\begin{align} \label{eq:fac-tQ-orig}
\tilde{Q}_{\gamma^z} \bigl(\zeta, {\mu}^2 z^2\bigr)
=&  \sum_{n} C_n ({\mu}^2 z^2)  \frac{(-i \zeta)^n}{n!} a_{n + 1}
 \left(\mu\right)
 \\
= & \sum_{n} C_n (\mu^2z^2)
\frac{(-i \zeta)^n}{n!}  \int_{-1}^1 dy\,y^n q \left( y, \mu\right)
 \,.  \nn
\end{align}
It should be noted that the only approximation we have made so far is ignoring
the higher-twist effects that are suppressed by small $z^2$ and the large momentum $P^z$ of the nucleon. In the limit of $P_z^2\gg M^2,\Lambda_{\rm QCD}^2$, we have $P^0\sim P^z$, so even if $\mu_0=0$ in Eq.~(\ref{eq:ope-tq}), the leading approximation of $\tilde{O}_{\gamma^0}(z)$ is still given by the twist-2 contributions in Eq.~(\ref{eq:fac-tQ-orig}), just with modified coefficients $C_n$.

Based on the OPE results in Eq.~(\ref{eq:fac-tQ-orig}), we can  derive a  factorization formula for the Euclidean \spcorr. First of all, let us
define a function $\mathcal{C} (\alpha, {\mu}^2 z^2)$:
\begin{align}  \label{eq:fac-C}
\mathcal{C} (\alpha, {\mu}^2 z^2) \equiv \int\! \frac{d \zeta}{2 \pi} \,
 e^{i \alpha \zeta}  \sum_n C_n ({\mu}^2 z^2)  \frac{(-i \zeta)^n}{n!}
 \,.
\end{align}
From \eq{fac-tQ-orig} and the renormalized analog of the Fourier-transform relation in \eq{pseudo} $\mathcal{C}(\alpha,\mu^2z^2)$ corresponds to a pseudo-PDF in the special case where $a_{n+1}(\mu)=1$. The analysis of Refs.~\cite{Radyushkin:1983wh,Radyushkin:2016hsy} implies that the support of $\mathcal{C}(\alpha,\mu^2z^2)$ is $-1 \le \alpha \le 1$. Noting that
\begin{align}
  \int d\alpha\, e^{-i\alpha(y \zeta)}\, \mathcal{C} (\alpha, {\mu}^2 z^2)
   &= \sum_n C_n(\mu^2 z^2) \frac{(-i\zeta y)^n}{n!} \,,
\end{align}
we find from \eq{fac-tQ-orig} that
\begin{align}
\tilde{Q}(\zeta,\mu^2 z^2)
= &\int_{-1}^1\!\! dy\int_{-1}^1\!\! d\alpha\ e^{-i\alpha(y\zeta)}\, \mathcal{C}(\alpha,\mu^2z^2)\, q(y,\mu) \,.
\end{align}
Finally, using the inverse transform of the renormalized analog of \eq{Qdefn},
\begin{align} \label{eq:FTq}
Q(\zeta,\mu) &= \int_{-1}^1\!\! dy\: e^{-i y\zeta}\,
  q(y,\mu) \,.
\end{align}
we obtain
\begin{align} \label{eq:io-Q-fact}
\tilde{Q}(\zeta,\mu^2 z^2)
=& \int_{-1}^1 d\alpha\ \mathcal{C}(\alpha,\mu^2z^2)\, Q(\alpha \zeta,\mu)
 + {\cal O}(z^2\Lambda_{\rm QCD}^2)
\,.
\end{align}
The result in \eq{io-Q-fact} is the factorization formula for the lattice calculable \spcorr $\tilde Q(\zeta,\mu^2 z^2)$ which expresses it in terms of the light-cone correlation $Q(\zeta,\mu)$ that defines the PDF. It has the same structure as the factorization formula for the \spcorr used for the calculation of the pion distribution amplitude in Ref.~\cite{Braun:2007wv,Bali:2017gfr}.

\subsection{Factorization for the quasi-PDF}

The renormalized quasi-PDF is defined as a Fourier transform of the renormalized \spcorr,
\begin{align} \label{eq:fac-tq-orig0}
\tilde{q} \Bigl(  x, \frac{\mu}{P_z}\Bigr)
\equiv &\int \frac{d \zeta}{2 \pi}\: e^{ix \zeta}\:  \tilde{Q} \biggl(\zeta, \frac{{\mu}^2\zeta^2}{P_z^2}\biggr) \,.
\end{align}
Note that we could use either $\tilde{Q}_{\gamma^z}$ or $\tilde{Q}_{\gamma^0}$ here.
Using the result for the \spcorr in Eq.~(\ref{eq:fac-tQ-orig}) this gives
\begin{align} \label{eq:fac-tq-orig}
\tilde{q} \Bigl( & x, \frac{\mu}{P_z}\Bigr)
 \\
= &  \int_{-1}^1 dy \biggl[ \int \frac{d \zeta}{2 \pi} e^{ix \zeta}
\sum_{n=0} C_n \Bigl( \frac{{\mu}^2\zeta^2}{P_z^2}\Bigr)  \frac{(-i
	\zeta)^n}{n!} y^n \biggr] q \left( y, \mu \right)
\nn\\
= &  \int_{-1}^1 {dy\over |y|} \biggl[\int \frac{d \zeta}{2 \pi} e^{i
	\frac{x}{y} \zeta}  \sum_{n=0} C_n \Bigl( \frac{{\mu}^2\zeta^2 }{(yP^z)^2}\Bigr)\,  \frac{(-i \zeta)^n}{n!}  \biggr] q \left( y,\mu \right) \,.
  \nn
\end{align}
Already, one can see that the matching kernel is a function of $x / y$
and $\mu / (|y|P^z)$. We define the kernel as
\begin{align}   \label{eq:fac-c}
C \left( {x\over y}, \frac{\mu}{|y|P^z} \right)
   \equiv \int\! \frac{d \zeta}{2 \pi} \: e^{i {x\over y} \zeta}
   \sum_{n=0} C_n \Bigl(
\frac{{\mu}^2\zeta^2}{(yP^z)^2}\Bigr) \, \frac{(-i \zeta)^n}{n!}
 \,,
\end{align}
and then Eq.~(\ref{eq:fac-tq-orig}) can be rewritten as
\begin{align} \label{eq:fac-tq}
\tilde{q} \left( x, \frac{\mu}{P^z}\right)
  = \int_{-1}^1 \frac{dy}{|y|}\: C \Bigl(
   \frac{x}{y}, \frac{\mu}{|y|P^z} \Bigr)\: q \left( y,\mu\right)
   \,,
\end{align}
which is the $\overline{\rm MS}$ factorization formula for the quasi-PDF. This result shows that the factorization formula in Eq.~(\ref{eq:momfact}) must have $p^z=|y|P^z$ for the quasi-PDF in the $\overline{\rm MS}$ scheme. We will show that this remains true for any quasi-PDF renormalization scheme in \sec{ren}. This differs
from the choice $p^z=P^z$ which had been conjectured and used in the early papers on the quasi-PDF~\cite{Ji:2013dva,Ji:2014gla,Xiong:2013bka}. Physically the correct result in \eq{fac-tq} can be understood as the fact that the matching coefficient is only sensitive to the perturbative partonic dynamics, and hence it is the magnitude of the partonic momentum $|y| P^z$ which appears, rather than the hadronic momentum $P^z$.

Taking the moment of the quasi-PDF using \eq{fac-tq-orig0} gives
\begin{align} \label{eq:momentqpdf}
 &\int_0^1\!\! dx\: x^n \tilde{q} \left( x, \frac{\mu}{P^z}\right)
= \Big(i\frac{d}{d\zeta}\Big)^n \tilde{Q} \biggl(\zeta, \frac{{\mu}^2\zeta^2}{P_z^2}\biggr) \Bigg|_{\zeta\to 0}
 \nn\\
& \quad =  \sum_{n'}  \Big(i\frac{d}{d\zeta}\Big)^n \bigg[ C_{n'} \Big(\frac{{\mu}^2 \zeta^2}{P_z^2}\Big)  \frac{(-i \zeta)^{n'}}{n'!} \bigg] \Bigg|_{\zeta\to 0}  a_{n' + 1}
\left(\mu\right) \,.
\end{align}
Since the $C_{n'}$ coefficients have $\ln(\zeta^2)$ dependence, the derivative for $n'=n$ will always have a logarithmic singularity as $\zeta\to 0$, and there will be even more singular terms for $n'<n$. This explains why the short distance Wilson coefficient causes the moments not to exist for the quasi-PDF.

\subsection{Factorization for the pseudo-PDF}

The renormalized pseudo-PDF is the Fourier transform of the renormalized \spcorr
\begin{align} \label{eq:pseudoRen}
\mathcal{P} \left( x, \mu^2 z^2\right)
   = \int_{-\infty}^\infty \! \frac{d \zeta}{2\pi}\:
    e^{ix \zeta}\: \tilde{Q}\bigl( \zeta, \mu^2 z^2 \bigr)
   \,.
\end{align}
Since both the pseudo-PDF and \spcorr are multiplicatively renormalized in a $\zeta$-independent manner, this follows immediately from \eq{pseudo}.
If we take \eq{io-Q-fact} and Fourier transform the \spcorr $\tilde Q(\zeta,\mu^2z^2)$ into the pseudo-PDF, and light-cone correlation $Q(\alpha\zeta,\mu)$ into the PDF, then we immediately obtain the factorization formula for the pseudo-PDF,
\begin{align}\label{eq:ps-q-fact}
\mathcal{P}(x,z^2\mu^2) =
 &\int_{|x|}^1 {dy\over |y|}\
  \mathcal{C}\left({x\over y},\mu^2 z^2\right) q(y,\mu)\nn\\
&+\int^{-|x|}_{-1} {dy\over |y|}\
  \mathcal{C}\left( \frac{x}{y}, \mu^2z^2\right) q(y,\mu)
  \nn \\
 &+ {\cal O}(z^2\Lambda_{\rm QCD}^2)\,,
\end{align}
which is the small $z^2$ factorization formula in Eq.~(\ref{eq:pseudofact}). The upper and lower limits of the integrals in \eq{ps-q-fact} follow immediately from the support $-1\le \alpha \le 1$ of the matching coefficient $\mathcal{C}(\alpha,z^2\mu^2)$, and we recall that we also have $-1\le x\le 1$ for the pseudo-PDF on the LHS.

Since the range of $x$ is bounded for the pseudo-PDF the terms in the series expansion of the exponential in \eq{pseudoRen} exist,
\begin{align}
  \tilde Q(\zeta,\mu^2 z^2)
   = \sum_{n=0}^\infty \int_{-1}^{1}\!\! dx\: \frac{(-i\zeta)^n x^n}{n!}
     {\cal P}(x,\mu^2 z^2) \,.
\end{align}
Comparing with \eq{fac-tQ-orig} this implies that the moments of the pseudo-PDF are given by
\begin{align}  \label{eq:Pmoment}
  \int_{-1}^{1}\!\! dx\: x^n \:  {\cal P}(x,\mu^2 z^2)
  &= C_n(\mu^2 z^2)\, a_{n+1}(\mu) \,.
\end{align}

So far we have proven the large $P^z$ factorization of the quasi-PDF and small $z^2$ factorization of the spatial correlation and pseudo-PDFs. After
deriving one factorization, it immediately leads to the others, since they are just different representations of the same spatial correlator. Indeed, we see that the quasi-PDF and pseudo-PDF are related at leading power by their definitions:
\beq \label{eq:equiv}
\tilde{q}\Bigl(x,{\mu\over P^z}\Bigr)
  = \int\! {d\zeta\over 2\pi}\ e^{ix\zeta}\int_{-1}^1\!\! dy \ e^{-iy\zeta}
  \: \mathcal{P}\biggl(y,{\mu^2 \zeta^2\over P_z^2}\biggr)\,,\nn\\
\eeq
where we have used $z=\zeta/P^z$.
Based on Eq.~(\ref{eq:fac-C}) and Eq.~(\ref{eq:fac-c}), the Wilson coefficients in their factorization theorems also maintain the same relationship,
\begin{align} \label{eq:ftc-C}
 C\left(\xi,{\mu\over |y|P^z}\right)
  &= \int\! {d\zeta\over 2\pi}\ e^{i\xi\zeta}\int_{-1}^1\!\! d\alpha\: e^{-i\alpha\zeta}\: \mathcal{C}\left(\alpha,{\mu^2 \zeta^2\over (yP^z)^2}\right) .
\end{align}
For the relations in \eqs{equiv}{ftc-C} the same choice of $\Gamma=\gamma^0$ or $\gamma^z$ should be used in the quasi- and pseudo-PDFs, or their corresponding coefficients.

In summary, there is a unique factorization formula that matches the quasi-PDF, \spcorr and pseudo-PDF to the PDF. Since their factorizations into the PDF all require small distances and have large nucleon momentum, the setup for their lattice calculations must also be the same. Therefore, the LaMET and pseudo distribution approaches are in principle equivalent to each other, and they differ perhaps only by effects related to their implementation on the lattice.

In Ref.~\cite{Radyushkin:2017cyf} it was speculated that one can study a ratio function
\beq \label{eq:ratio}
\tilde{Q}(\zeta,z^2,a^{-1})/\tilde{Q}(0,z^2,a^{-1})
\eeq
on a lattice with spacing $a$, and the $O(z^2)$ corrections may cancel approximately. This idea was tested in Ref.~\cite{Orginos:2017kos} in lattice QCD, and the results show that the ratio evolves slowly in $z^2$ at small values. It is then interesting to consider what type of non-perturbative information can be extracted from this ratio.

This question can be answered using the small $z^2$ factorization for the \spcorr.
 According to Eq.~(\ref{eq:fac-tQ-orig}),
\beq
	\tilde{Q}(0,\mu^2 z^2) = C_0(\mu^2z^2) + O(z^2\Lambda_{\rm QCD}^2)\,,
\eeq
where in the $\overline{\rm MS}$ scheme to one-loop
\beq  \label{eq:C0}
	C_0(\mu^2z^2) = 1 + {\alpha_sC_F\over 2\pi} \biggl[ {3\over2} \ln(\mu^2z^2 e^{2\gamma_E}/4)+{5\over2} \biggr] \,,
\eeq
which was also derived recently in \cite{Zhang:2018ggy}.
Then the ratio becomes
\begin{align} \label{eq:Qtratio}
	\frac{\tilde{Q}\left(\zeta, {\mu}^2 z^2\right)}{\tilde{Q}\left(0, {\mu}^2 z^2\right)}
	=&\sum_n \frac{C_n (\mu^2z^2)}{C_0 (\mu^2z^2)}	\frac{(-i \zeta)^n}{n!} a_{n+1}(\mu)\nn\\
	&+ O(z^2\Lambda_{\rm QCD}^2)\,.
\end{align}
Using \eq{Pmoment} and our $\overline{\rm MS}$ one-loop perturbative pseudo-PDF result in \eq{1looppseudopdfb} below we find for $\Gamma=\gamma^0$ that
\begin{align}
  C_n(\mu^2 z^2) &= 1 +\frac{\alpha_s C_F}{2\pi} \biggl[ \Bigl(
  \frac{3\!+\!2n}{2\!+\!3n\!+\!n^2}
  \!+\! 2H_n\Bigr) \ln\frac{\mu^2z^2 e^{2\gamma_E}}{4}
  \nn\\
 &\quad\quad
  +   \frac{5\!+\!2n}{2\!+\!3n\!+\!n^2}+2(1-H_n)H_n - 2 H_n^{(2)}
  \biggr] ,
\end{align}
where the Harmonic numbers are $H_n=\sum_{i=1}^n 1/i$ and $H_n^{(2)} =\sum_{i=1}^n 1/i^2$.  For the case $\Gamma=\gamma^z$ we have $C_n^{\gamma^z}(\mu^2 z^2) = C_n(\mu^2 z^2) + \Delta C_n^{\gamma^z}(\mu^2 z^2)$ with
\begin{align}
  \Delta C_n^{\gamma^z}(\mu^2 z^2)
   &= \frac{\alpha_s C_F}{2\pi} \: \frac{2}{2\!+\!3n\!+\!n^2} \,,
\end{align}
which also modifies \eq{C0} for $n=0$.
At small $z^2$ where the perturbative expansion with $\mu\simeq 1/|z|$ is valid, the ratio in \eq{Qtratio} has a weak logarithmic dependence on $|z|$, which is consistent with the lattice findings in Refs.~\cite{Orginos:2017kos,Karpie:2017bzm}. The weak dependence on $|z|$ can be quantitatively described by an evolution equation in $\ln z^2$~\cite{Radyushkin:2017cyf,Radyushkin:2017lvu}. According to our OPE analysis, here $\tilde{Q}(0,\mu^2 z^2)$ only serves as an overall normalization factor which is contaminated by higher-twist corrections, and the $\ln z^2$ evolution can be put in accurate terms with the factorization formula in Eq.~(\ref{eq:io-Q-fact}) that enables us extract the PDF from the ratio function. The same point was demonstrated by work done very recently in~\cite{Zhang:2018ggy}, which appeared simultaneously with our paper.

\section{Equivalence at one-loop order}
\label{sec:equiv}

As has been proven in Sec.~\ref{sec:ope}, the quasi-PDF and pseudo-PDF as well as their matching coefficients are related by a simple Fourier transform in Eq.~(\ref{eq:equiv}) and Eq.~(\ref{eq:ftc-C}). This relation is valid to all orders in perturbation theory. In this section we check the relations in Eq.~(\ref{eq:equiv}) and Eq.~(\ref{eq:ftc-C}) at one-loop order. We choose $\Gamma=\gamma^0$ for our main presentation, but also quote final results for the case $\Gamma=\gamma^z$.

In the Feynman gauge, we calculate the quark matrix elements of the unpolarized iso-vector quasi-PDF, pseudo-PDF, and light-cone PDF at one-loop order in dimensional
regularization with $d=4-2\epsilon$. The external quark state is chosen to be on-shell and massless, and we regularize the UV and collinear divergences by $1/\epsilon_{\mbox{\tiny UV}}\ (\epsilon_{\mbox{\tiny UV}}>0)$ and $1/\epsilon_{\mbox{\tiny IR}}\ (\epsilon_{\mbox{\tiny IR}}<0)$ respectively. The one-loop order Feynman diagrams are shown in Fig.~\ref{fig:diagram}.

\begin{widetext}

\begin{figure*}[t!]
	\centering
	\includegraphics[width=.9\textwidth]{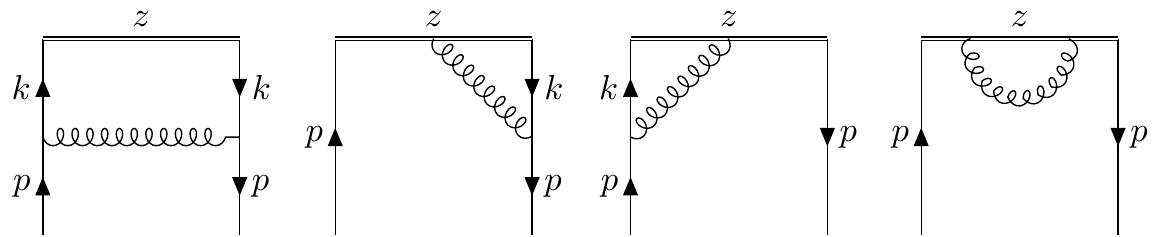}
	\caption{One-loop Feynman diagrams for the quasi-PDF, \spcorr and pseudo-PDFs. The first one is named ``vertex", the second and third ones are named ``sail", and the last one ``tadpole". The standard quark self energy wavefunction is also included.}
	\label{fig:diagram}
\end{figure*}

In an on-shell quark state with momentum $p^\mu=(p^0=p^z,0,0,p^z)$, for $\Gamma=\gamma^0$, each diagram gives
\begin{align}  \label{eq:qtloopint1}
\tilde{Q}^{(1)}_\text{vertex}(\zeta,z^2,\epsilon)
&= {\mu^{2\epsilon} \iota^\epsilon \over 2p^0}\bar{u}(p)\int {d^d k\over (2\pi)^d}(-ig T^a\gamma^\mu) {i\over \slashed k} \gamma^0 {i\over\slashed k} (-i gT^a\gamma^\nu){ - i g_{\mu\nu} \over (p-k)^2}u(p)e^{-i k^z z}\ ,
  \\
\tilde{Q}^{(1)}_\text{sail}(\zeta,z^2,\epsilon)
&= {\mu^{2\epsilon} \iota^\epsilon\over 2p^0}\bar{u}(p)\int {d^d k\over (2\pi)^d} (ig T^a \gamma^0) {1\over i(p^z-k^z)} \left( e^{-ip^z z}- e^{-ik^z z}\right)  \delta^{\mu z} {i\over \slashed k} (-ig T^a \gamma^\nu) {-ig_{\mu\nu}\over (p-k)^2}u(p)
  \nonumber\\
&+ {\mu^{2\epsilon}\iota^\epsilon \over 2p^0}\bar{u}(p) \int {d^d k\over (2\pi)^d} (-ig T^a \gamma^\nu){i\over \slashed k} (ig T^a \gamma^0) {1\over i(p^z-k^z)} \left( e^{-ip^z z}- e^{-ik^z z}\right)  \delta^{\mu z}   {-ig_{\mu\nu}\over (p-k)^2}u(p)\ ,
  \nn \\
\tilde{Q}^{(1)}_\text{tadpole}(\zeta,z^2,\epsilon)
&= {\mu^{2\epsilon}\iota^\epsilon \over 2p^0}\bar{u}(p)\int {d^d k\over (2\pi)^d} (-g^2)C_F \gamma^0 \delta^{\mu z}\delta^{\nu z}\left({e^{-ip^z z}- e^{-ik^z z}\over (p^z-k^z)^2} - {ze^{-ip^z z}\over i(p^z-k^z)}\right){-ig_{\mu\nu}\over (p-k)^2}u(p)\ ,
  \nn
\end{align}
where $\iota = e^{\gamma_E}/(4\pi)$ is included to implement $\mu$ in the $\overline{\rm MS}$ scheme, $C_F=4/3$, and $T^a$ is the SU(3) color matrix in the fundamental representation. The second term in the brackets in the last line, which is proportional $z$, does not contribute to the loop integral as it is odd under the exchange of $p^z-k^z\to -(p^z-k^z)$. The quark self-energy correction is $\tilde{Q}_{\rm w.fn.}^{(1)}(\zeta,z^2,\epsilon) =\delta Z_\psi\: \tilde{Q}^{(0)}(\zeta,z^2)$ with the tree level matrix element $\tilde{Q}^{(0)}(\zeta,z^2)=e^{-i\zeta}$ and on-shell renormalization constant $\delta Z_\psi$,
\beq
\delta Z_\psi = {\alpha_sC_F\over 2\pi}\left(-{1\over2}\right)\left({1\over\epsilon_{\mbox{\tiny UV}}} - {1\over \epsilon_{\mbox{\tiny IR}}}\right)\ .
\eeq

After carrying out the loop integrals in Eq.~(\ref{eq:qtloopint1}) according to the method in Ref.~\cite{Ji:2017rah}, we obtain
\begin{align}\label{eq:1loopdiagram}
\tilde{Q}^{(1)}_{\rm vertex}(\zeta,z^2,\epsilon)
=\,& {\alpha_sC_F\over 2\pi}e^{\epsilon\gamma_E} \int_0^1 du\ (1-\epsilon)(1-u) e^{-iu \zeta}\Gamma(-\epsilon) 4^{-\epsilon} \big(\mu |z|\big)^{2\epsilon}
  \\
 =\,& {\alpha_sC_F\over 2\pi} e^{\epsilon\gamma_E}\biggl({\mu|z|\over2}\biggr)^{2\epsilon}
 \frac{(-1)\Gamma(2-\epsilon)}{\epsilon_{\mbox{\tiny IR}}}
 {1-i\zeta - e^{-i\zeta}\over \zeta^2}
  \,, \nonumber\\
\tilde{Q}^{(1)}_{\rm sail}(\zeta,z^2,\epsilon)
=\, & {\alpha_sC_F\over 2\pi}  e^{\epsilon\gamma_E}\left[(i\zeta) \int_0^1 du \int_0^1 dt (2-u) e^{-i(1-ut)\zeta} \Gamma(-\epsilon) 4^{-\epsilon}(t^2z^2\mu^2)^{\epsilon}\right.
  \nonumber\\
 &\left. -\int_0^1 du \ e^{-iu \zeta}\Gamma(-\epsilon) 4^{-\epsilon}(z^2\mu^2)^{\epsilon}  + \left({1\over\epsilon_{\mbox{\tiny UV}}} - {1\over \epsilon_{\mbox{\tiny IR}}}\right) e^{-i\zeta}\right]
  \nonumber\\
 =\,& {\alpha_sC_F\over 2\pi} e^{\epsilon\gamma_E} \left({\mu|z|\over2}\right)^{2\epsilon}
 \left\{\frac{-\Gamma(1-\epsilon)}{\epsilon_{\mbox{\tiny IR}}}
 {2(1-\epsilon)\over 1-2\epsilon} \left[{1-e^{-i\zeta}\over -i\zeta} + e^{-i\zeta}(-i\zeta)^{-2\epsilon}\big(\Gamma(2\epsilon)-\Gamma(2\epsilon,-i\zeta)\big)
 \right]\right.
  \nn\\
 &\left. + {\Gamma(1-\epsilon)\over \epsilon_{\mbox{\tiny IR}} } \frac{e^{-i\zeta}}{\epsilon} + \left({1\over\epsilon_{\mbox{\tiny UV}}} - {1\over \epsilon_{\mbox{\tiny IR}}}\right) e^{-i\zeta} \right\}
  \,, \nn\\
\tilde{Q}^{(1)}_{\rm tadpole}(\zeta,z^2,\epsilon)
=\, & {\alpha_sC_F\over 2\pi} e^{\epsilon\gamma_E}\left({\mu|z|\over2}\right)^{2\epsilon}  {\Gamma(1-\epsilon)\over \epsilon_{\mbox{\tiny UV}} (1-2\epsilon)}e^{-i\zeta}
 \ . \nn
\end{align}
For simplicity we have left out the tree level multiplicative spinor factor $\bar u\gamma^0 u$ when quoting one-loop results in \eq{1loopdiagram}, and will continue to do so for the \spcorr, quasi-PDF, and pseudo-PDF results quoted below.
Since $\alpha_s^{\rm bare}=\alpha_s(\mu) \mu^{2\epsilon} Z_g^2 = \alpha_s(\mu) \mu^{2\epsilon} + {\cal O}(\alpha_s^2)$ is $\mu$-independent we do not include $\mu$ as an argument in the bare functions.
In the final result for each term we have also specified whether $1/\epsilon$ factors (that remain after expanding about $\epsilon\to 0$) are IR or UV divergences. Combined with the wavefunction corection, the bare \spcorr $\tilde{Q}^{(1)}(\zeta,z^2,\epsilon)$ is
\begin{align} \label{eq:1loopioffe}
\tilde{Q}^{(1)}(\zeta,z^2,\epsilon)
&= {\alpha_sC_F\over 2\pi}e^{\epsilon\gamma_E} \left({\mu|z|\over2}\right)^{2\epsilon} \Bigg\{
 {3\over2}\left({1\over \epsilon_{\mbox{\tiny UV}}}-{1\over \epsilon_{\mbox{\tiny IR}}}\right)e^{-i\zeta}
 +  \frac{(-1)\Gamma(2-\epsilon)}{\epsilon_{\mbox{\tiny IR}}}
  \left[ {(1-i\zeta - e^{-i\zeta})\over \zeta^2}
 \right.
 \\
& \qquad\qquad\qquad\qquad\qquad\quad \left.
 +{2\over 1-2\epsilon}\left\{ {1-e^{-i\zeta}\over -i\zeta} + e^{-i\zeta}(-i\zeta)^{-2\epsilon}
 \big(\Gamma(2\epsilon)-\Gamma(2\epsilon,-i\zeta)\big) - \frac{e^{-i\zeta}}{2\epsilon} \right\} \right]
 \Bigg\}\ . \nn
\end{align}
Note that as $\epsilon\to 0$ the terms in the innermost curly brackets have no $1/\epsilon$ term.  Also we can verify that in the local limit of the operator that the bare one-loop correction vanishes as expected by conservation of the vector current:
\begin{align} \label{eq:Qvcc}
\lim_{z\to 0} \tilde{Q}^{(1)}(z P^z,z^2,\epsilon)
  = \lim_{z\to 0}  {\alpha_sC_F\over 2\pi}e^{\epsilon\gamma_E} \left({\mu|z|\over2}\right)^{2\epsilon} \bigg\{
 {3\over2}\left({1\over \epsilon_{\mbox{\tiny UV}}}-{1\over \epsilon_{\mbox{\tiny IR}}}\right)
  + \frac{3}{2\epsilon_{\mbox{\tiny IR}}} \bigg\} = 0
 \,,
\end{align}
where we note that it is important that the $1/\epsilon_{\mbox{\tiny IR}}$ terms cancel since the assumption $\epsilon>0$ is only valid for the $1/\epsilon_{\mbox{\tiny UV}}$ term.
The corresponding bare pseudo-PDF is
\begin{align}\label{eq:1looppseudo}
{\cal P}^{(1)}(x,z^2,\epsilon)
 &= {\alpha_sC_F\over 2\pi} \left[-1+x + \frac{2}{(1-2\epsilon)} -
 \left( {2\over (1-2\epsilon)}{1\over (1-x)^{1-2\epsilon}}\right)^{[0,1]}_{+(1)}\right]e^{\epsilon\gamma_E}\left({\mu|z|\over2}\right)^{2\epsilon} \frac{\Gamma(2-\epsilon)}{\epsilon_{\mbox{\tiny IR}}} \theta(x)\theta(1-x)\nonumber\\
 &\quad +{\alpha_sC_F\over 2\pi} \:
  e^{\epsilon\gamma_E}\left({\mu|z|\over2}\right)^{2\epsilon}{3\over2}\left({1\over \epsilon_{\mbox{\tiny UV}}}-{1\over \epsilon_{\mbox{\tiny IR}}}\right)\, \delta(1-x)
\ .
\end{align}
Since we will encounter plus functions over different domains below, we define a plus function at $x=x_0$ within a given domain $D$ so that
\beq
	\int_D dx\ \big[ g(x)\big]_{+(x_0)}^D\, h(x) = \int_D dx\ g(x) \left[ h(x) - h(x_0)\right]\,.
\eeq
(See \app{ep} for more details.)
It is straightforward to confirm that the bare pseudo-PDF satisfies the local vector current conservation, $\lim_{z\to 0} \int\! dx \, {\cal P}^{(1)}(x,z^2\mu^2,\epsilon) =0$, with the same cancellation as in \eq{Qvcc}. 	
	
Now, according to the relations between the quasi-PDF and the \spcorr or pseudo-PDFs in Eqs.~(\ref{eq:fac-tq-orig0},\ref{eq:equiv}), we can do a Fourier or double Fourier transform of the results in Eqs.~(\ref{eq:1loopioffe},\ref{eq:1looppseudo}) to get the quasi-PDF. Despite its straightforwardness, the Fourier transform is subtle and the details are provided in \app{ft}. Here we simply quote the result for the bare quasi-PDF,
\begin{align}  \label{eq:1loopquasi}
\tilde{q}^{(1)}(x,p^z,\epsilon)
	=& {\alpha_sC_F\over 2\pi}\left\{{3\over2}\left({1\over \epsilon_{\mbox{\tiny UV}}}-{1\over \epsilon_{\mbox{\tiny IR}}}\right)\delta(1-x) + {\Gamma(\epsilon+{1\over2})e^{\epsilon\gamma_E}\over \sqrt{\pi}}{\mu^{2\epsilon}\over p_z^{2\epsilon}}{1-\epsilon \over \epsilon_{\mbox{\tiny IR}} (1-2\epsilon)} \right.\nn\\
	&\times\left.\left[|x|^{-1-2\epsilon}\left(1+x+{x\over2}(x-1+2\epsilon)\right)-  |1-x|^{-1-2\epsilon}\left(x+{1\over2}(1-x)^2\right) + I_3(x)\right] \right\}\,,
\end{align}
where
\beq
	I_3(x) = \theta(x-1)\left(x^{-1-2\epsilon}\over x-1\right)^{[1,\infty]}_{+(1)} -\theta(x)\theta(1-x) \left(x^{-1-2\epsilon}\over 1-x\right)^{[0,1]}_{+(1)} - \delta(1-x) \pi\csc(2\pi\epsilon) + \theta(-x) {|x|^{-1-2\epsilon}\over x-1}\,.
\eeq
After some algebra one can confirm that the bare quasi-PDF satisfies local vector current conservation, with $\int\! dx\, \tilde{q}^{(1)}(x,p^z,\epsilon) = 0$. To verify this result one must carefully separate out $1/\epsilon_{\mbox{\tiny UV}}$ factors arising from requiring $\epsilon>0$ to obtain convergence at $x=\pm \infty$, and $1/\epsilon_{\mbox{\tiny IR}}$ factors that arise from requiring $\epsilon<0$ to obtain convergence at $x=1$.
	
An alternate method of obtaining the quasi-PDF is to directly calculate it from the Feynman diagrams by first Fourier transforming $z$ into $xp^z$. As a result, the factors $(e^{-ip^z z}- e^{-ik^z z})$ are transformed into $[\delta(p^z-xp^z)-\delta(k^z-xp^z)]$, and all the loop integrals reduce to $(d-1)$-dimensional ones. This is the procedure for the matching calculations of the quasi-PDF used in Refs.~\cite{Xiong:2013bka,Stewart:2017tvs,Wang:2017qyg}, and is distinct from doing the Fourier transformation after fully carrying out the integrals as in Eqs.~(\ref{eq:1loopdiagram}--\ref{eq:1loopquasi}). As a cross-check we have confirmed in \app{quasi} that we obtain the exact same bare quasi-PDF in \eq{1loopquasi} from both procedures.

Now we consider the $\epsilon$ expansion to obtain $\overline{\rm MS}$ renormalized results for the \spcorr, pseudo-PDF, and quasi-PDF. Expanding the \spcorr in $\epsilon$ we obtain
\begin{align}
 \tilde{Q}^{(1)}(\zeta,z^2,\epsilon)=\delta \tilde{Q}^{(1)}(\zeta,z^2,\mu,\epsilon_{\mbox{\tiny UV}})+\tilde{Q}^{(1)}(\zeta,z^2,\mu,\epsilon_{\mbox{\tiny IR}})+{\cal O}(\epsilon)
\end{align}
with the $\overline{\rm MS}$ counterterm and renormalized \spcorr given by
\begin{align} \label{eq:1loopiofferen}
\delta \tilde{Q}^{(1)}(\zeta,z^2,\mu,\epsilon_{\mbox{\tiny UV}})
&= {\alpha_sC_F\over 2\pi}
  e^{-i\zeta} {3\over2}  {1\over \epsilon_{\mbox{\tiny UV}}}
 \,, \\
\tilde{Q}^{(1)}(\zeta,z^2,\mu,\epsilon_{\mbox{\tiny IR}})
&= {\alpha_sC_F\over 2\pi} \Bigg\{
  \frac{3}{2} \Big(\ln{\mu^2 z^2 e^{2\gamma_E} \over 4} +1\Big) e^{-i\zeta}
  + \Bigl( -\frac{1}{ \epsilon_{\mbox{\tiny IR}} } - \ln\frac{z^2\mu^2e^{2\gamma_E}}{4} - 1 \Bigr) h(\zeta)
  + \frac{2(1 \!-\!i\zeta\!-\! e^{-i\zeta})}{\zeta^2}
  \nn\\
&\qquad\qquad\quad
  + 4 i \zeta e^{-i\zeta}\, {}_3F_3(1,1,1,2,2,2,i\zeta)
  \Bigg\}
  \,. \nn
\end{align}
Here ${}_3F_3$ is a hypergeometric function and the Fourier transform of $\big[(1+x^2)/(1-x)\big]_{+(1)}^{[0,1]}$ gives the function
\begin{align}
  h(\zeta) &=\frac{3}{2}\, e^{-i\zeta}
  + \frac{1+i\zeta-e^{-i\zeta}-2i\zeta e^{-i\zeta}}{\zeta^2}
  - 2 e^{-i\zeta} \big[ \Gamma(0,-i\zeta)\!+\!\gamma_E\!+\!\ln(-i\zeta) \big]
  \,.
\end{align}
For the position space PDF we have
\begin{align}
  Q^{(1)}(\zeta,\mu,\epsilon_{\mbox{\tiny IR}})
  &= -  {\alpha_sC_F\over 2\pi}  \frac{1}{  \epsilon_{\mbox{\tiny IR}} }  h(\zeta) \,.
\end{align}

Next we expand the bare pseudo-PDF from Eq.~(\ref{eq:1looppseudo}) in $\epsilon$ to obtain the $\overline{\rm MS}$ counterterm and renormalized pseudo-PDF as ${\cal P}^{(1)} (x,z^2,\epsilon)= \delta {\cal P}^{(1)} (x,z^2\mu^2,\epsilon_{\mbox{\tiny UV}})+{\cal P}^{(1)} (x,z^2\mu^2,\epsilon_{\mbox{\tiny IR}})+{\cal O}(\epsilon)$ with
\begin{align}\label{eq:1looppseudopdfb}
\delta {\cal P}^{(1)} &(x,z^2\mu^2,\epsilon_{\mbox{\tiny UV}})
	= {\alpha_sC_F\over 2\pi}\:
    {3\over 2\epsilon_{\mbox{\tiny UV}}} \: \delta(1-x)
  \,, \nn\\
{\cal P}^{(1)} &(x,z^2\mu^2,\epsilon_{\mbox{\tiny IR}})
\nn\\
	=&{\alpha_sC_F\over 2\pi}\left\{\left({1+x^2\over 1-x}\right)^{[0,1]}_{+(1)}\left[- \left({1\over \epsilon_{\mbox{\tiny IR}}}+\ln{e^{2\gamma_E}\over 4}\right)-\ln(z^2 \mu^2)-1\right] - \left(4\ln(1-x)\over 1-x\right)^{[0,1]}_{+(1)}  +2(1-x)\right\}\theta(x)\theta(1-x)\nonumber\\
	&  +{\alpha_sC_F\over 2\pi}\left[{3\over2}\ln(z^2\mu^2) + {3\over2}\ln{e^{2\gamma_E}\over 4}+{3\over2}\right]\delta(1-x)
 \,.
\end{align}
Note that the renormalized $\overline{\rm MS}$ pseudo-PDF depends explicitly on $\mu^2$, and satisfies the relation to the renormalized $\overline{\rm MS}$ \spcorr given in Eq.~(\ref{eq:pseudoRen}).
It is also interesting to note that having expanded in $\epsilon$, local vector current conservation is no longer satisfied by the limit of the renormalized $\overline{\rm MS}$ pseudo-PDF, since
\begin{align} \label{eq:Ppdfconsrv}
\lim_{z\to 0} \int\! dx \, {\cal P}^{(1)} (x,z^2\mu^2,\epsilon_{\mbox{\tiny IR}}) \simeq (3\alpha_s C_F/4\pi) \lim_{z\to 0} \ln(z^2\mu^2)
\end{align}
gives a divergent result. The same divergence is present in the one-loop $\overline{\rm MS}$ renormalized \spcorr in \eq{1loopiofferen}.  Although this is the case in $\overline{\rm MS}$, it does not need to be the case in other renormalization schemes.

For the quasi-PDF there are two methods that we can consider for the renormalized calculation, either expanding the bare result in \eq{1loopquasi} and renormalizing in $(x,p^z)$ space, or following our preferred definition in \eq{fac-tq-orig0} and Fourier transforming the renormalized \spcorr in \eq{1loopiofferen}.  Although these two approaches will lead to the same final result for $C$ for practical applications, there is a subtle difference that we will explain.

First consider the renormalization of the quasi-PDF done in $(x,p^z)$ space. Expanding Eq.~(\ref{eq:1loopquasi}) in $\epsilon$, and writing $\tilde{q}^{(1)}(x,p^z,\epsilon) =\delta \tilde{q}^{(1)}(x,\mu/|p^z|,\epsilon_{\mbox{\tiny UV}})+\tilde{q}^{(1)}(x,\mu/|p^z|,\epsilon_{\mbox{\tiny IR}}) +{\cal O}(\epsilon)$ allows us to identify the $\overline{\rm MS}$ counterterm and renormalized quasi-PDF as
\begin{align}\label{eq:qPDFren}
\delta \tilde{q}^{(1)}(x,\mu/|p^z|,\epsilon_{\mbox{\tiny UV}})
=&\,  {\alpha_sC_F\over 2\pi} {3\over 2\epsilon_{\mbox{\tiny UV}}}
 \left[\delta(1-x) - {1\over2}{1\over x^2}\delta^+\Big({1\over x}\Big) - {1\over2}{1\over (1-x)^2}\delta^+\Big({1\over 1-x}\Big) \right]
 \,, \nn\\
\tilde{q}^{(1)}(x,\mu/|p^z|,\epsilon_{\mbox{\tiny IR}})
=&\, {\alpha_sC_F\over 2\pi}\left\{
\begin{array}{ll}
\displaystyle \left({1+x^2\over 1-x}\ln {x\over x-1} + 1 + {3\over 2x}\right)^{[1,\infty]}_{+(1)}- \left({3\over 2x}\right)^{[1,\infty]}_{+(\infty)} &\, x>1\nn\\[10pt]
\displaystyle \left({1+x^2\over 1-x}\left[- {1\over \epsilon_{\mbox{\tiny IR}}} - \ln{\mu^2\over 4p_z^2} + \ln\big(x(1-x)\big)\right] - {x(1+x)\over 1-x}\right)^{[0,1]}_{+(1)}  &\, 0<x<1\nn\\[10pt]
\displaystyle  \left(-{1+x^2\over 1-x}\ln {-x\over 1-x} - 1 + {3\over 2(1-x)}\right)^{[-\infty,0]}_{+(1)} - \left({3\over 2(1-x)}\right)^{[-\infty,0]}_{+(-\infty)} \quad &\, x<0
\end{array}\right.\nn\\[5pt]
& + {\alpha_sC_F\over 2\pi}\left[\delta(1-x) - {1\over2}{1\over x^2}\delta^+\Big({1\over x}\Big) - {1\over2}{1\over (1-x)^2}\delta^+\Big({1\over 1-x}\Big) \right] \left( {3\over2}\ln{\mu^2\over 4p_z^2} + {5\over2}\right)
 \,.
\end{align}
The details of working out the $\epsilon$ expansion of \eq{1loopquasi} are provided in \app{ep}, including definitions of the plus functions and $\delta$-functions at $x_0=\pm \infty$ that appear in the result quoted here. The $\overline{\rm MS}$ quasi-PDF obtained in \eq{qPDFren} still satisfies vector current conservation
\begin{align} \label{eq:qpdfconsrv}
 \int\! dx\: \tilde{q}^{(1)}(x,\mu/|p^z|,\epsilon_{\mbox{\tiny IR}})  =0 \,.
\end{align}
This is obviously the case for the plus function terms which individually integrate to zero, and is also true for the combination of $\delta$-functions which appears in \eq{qPDFren}.

The renormalized $\overline{\rm MS}$ quasi-PDF in Eq.~(\ref{eq:qPDFren}) differs slightly from that obtained using our definition in Eq.~(\ref{eq:fac-tq-orig0}). Using Eq.~(\ref{eq:fac-tq-orig0}) and the renormalized \spcorr in \eq{1loopiofferen} we instead obtain
\begin{align}\label{eq:qPDFft}
\delta \tilde{q}'^{(1)}(x,\mu/|p^z|,\epsilon_{\mbox{\tiny UV}})
=&\,  {\alpha_sC_F\over 2\pi} {3\over 2\epsilon_{\mbox{\tiny UV}}}\delta(1-x)\,, \\
\tilde{q}'^{(1)}(x,\mu/|p^z|,\epsilon_{\mbox{\tiny IR}})
=&\, {\alpha_sC_F\over 2\pi}\left\{
\begin{array}{ll}
\displaystyle \left({1+x^2\over 1-x}\ln {x\over x-1} + 1 + {3\over 2x}\right)^{[1,\infty]}_{+(1)}- \left({3\over 2x}\right)^{[1,\infty]}_{+(\infty)} &\, x>1\nn\\[10pt]
\displaystyle \left({1+x^2\over 1-x}\left[- {1\over \epsilon_{\mbox{\tiny IR}}} - \ln{\mu^2\over 4p_z^2} + \ln\big(x(1-x)\big)\right] - {x(1+x)\over 1-x}\right)^{[0,1]}_{+(1)}  &\, 0<x<1\nn\\[10pt]
\displaystyle  \left(-{1+x^2\over 1-x}\ln {-x\over 1-x} - 1 + {3\over 2(1-x)}\right)^{[-\infty,0]}_{+(1)} - \left({3\over 2(1-x)}\right)^{[-\infty,0]}_{+(-\infty)} \quad &\, x<0
\end{array}\right.\nn\\[5pt]
& + {\alpha_sC_F\over 2\pi}\left[\delta(1-x)\left( {3\over2}\ln{\mu^2\over 4p_z^2} + {5\over2}\right) +{3\over2}\gamma_E\left({1\over(x-1)^2}\delta^+({1\over x-1}) + {1\over(1-x)^2}\delta^+({1\over 1-x})\right)\right]
 \nn .
\end{align}
To carry out this calculation we defined the Fourier transformation of the singular function $\ln(\zeta^2)$ as
\begin{align} \label{eq:subtlety}
\int {d\zeta\over 2\pi}\ e^{i x \zeta} \ln\zeta^2
= & \biggl[ \left.{d\over d\eta} \int {d\zeta\over 2\pi}\ e^{i x \zeta} (\zeta^2)^\eta \biggr] \right|_{\eta=0}
=\biggl[ \left.{d\over d\eta} {4^\eta\over \Gamma(-\eta)} {\Gamma(\eta+1/2)\over \sqrt{\pi}} {[\theta(x)+\theta(-x)]\over |x|^{1+2\eta}} \biggr]
\right|_{\eta=0}\nn\\
=& \gamma_E \left[\left(-\delta(x) + {1\over x^2} \delta^+\Big({1\over x}\Big)\right) + \left(-\delta(x) + {1\over x^2} \delta^+\Big({1\over -x}\Big)\right)\right] \nn\\
& -\left[\left({1\over x}\right)_{+(0)}^{[0,1]} + \left({1\over x}\right)_{+(\infty)}^{[1,\infty]}\right]\theta(x) - \left[\left({1\over -x}\right)_{+(0)}^{[-1,0]} + \left({1\over -x}\right)_{+(\infty)}^{[-\infty,-1]}\right]\theta(-x)\,,
\end{align}
where we have used the results in Eqs.~(\ref{eq:ft},\ref{eq:expansion}) to derive the second and last equalities, and took the limit $\eta\to 0^+$ or $\eta\to 0^-$ when needed.  This $\tilde{q}'^{(1)}(x,\mu/|p^z|,\epsilon_{\mbox{\tiny IR}})$ does not satisfy vector-current conservation, and is different from $\tilde{q}^{(1)}(x,\mu/|p^z|,\epsilon_{\mbox{\tiny IR}})$ in \eq{qPDFren} only by the $\delta$-functions at $x_0=\pm\infty$. Within the function domain $-\infty < x < \infty$, they are exactly the same.  We will see below that both \eq{qPDFren} and \eq{qPDFft} eventually lead to the same result for the one-loop matching coefficient.

The final ingredient we need for the matching calculations is the PDF, whose one-loop bare matrix element can be written as a sum of an $\overline{\rm MS}$ counterterm and renormalized matrix element, $q^{(1)}(x,\epsilon) = \delta q^{(1)}(x,\epsilon_{\mbox{\tiny UV}})+q^{(1)}(x,\epsilon_{\mbox{\tiny IR}})$, where
\begin{align}
 \delta q^{(1)}(x,\epsilon_{\mbox{\tiny UV}}) &= 	{\alpha_sC_F\over 2\pi} {1\over \epsilon_{\mbox{\tiny UV}}} \left({1+x^2\over 1-x}\right)^{[0,1]}_{+(1)}  \theta(x)\theta(1-x)
  \,, \\
 q^{(1)}(x,\epsilon_{\mbox{\tiny IR}}) &= 	{\alpha_sC_F\over 2\pi} {(-1)\over \epsilon_{\mbox{\tiny IR}}} \left({1+x^2\over 1-x}\right)^{[0,1]}_{+(1)} \theta(x)\theta(1-x)
  \,. \nn
\end{align}
	
With the above results in hand we can now determine the matching coefficients up to one-loop order.  Using \eq{ps-q-fact} we find
\beq \label{eq:pseudoC}
{\cal C}^{(1)}(\alpha,z^2\mu^2)= {\cal P}^{(1)}(\alpha,z^2\mu^2,\epsilon_{\mbox{\tiny IR}}) - q^{(1)}(\alpha,\epsilon_{\mbox{\tiny IR}}) \,.
\eeq
Therefore the matching coefficient relating the pseudo-PDF and PDF in the $\overline{\rm MS}$ scheme with $\Gamma=\gamma^0$ is\footnote{A one-loop analysis of the \spcorr in the coordinate space also recently appeared in Refs.~\cite{Radyushkin:2017lvu,Radyushkin:2018cvn}. Our factorization result for the \spcorr in \eq{io-Q-fact} has a similar form to the hard part of the reduced \spcorr found in Eq.(3.35) of Ref.~\cite{Radyushkin:2017lvu} and Eq.(17) of Ref.~\cite{Radyushkin:2018cvn}. It is therefore interesting to compare our ${\cal C}(\alpha,z^2\mu^2)/C_0(\mu^2z^2)$ and this hard part. Our $\overline{\rm MS}$ result  \eq{ps-c} differs from Refs.~\cite{Radyushkin:2017lvu} due to the presence of the $2(1-\alpha)$ term. The result in the final version of Ref.~\cite{Radyushkin:2018cvn} agrees with ours. \eq{ps-c} also agrees with the original result derived in Ref.~\cite{Ji:2017rah}, up to the addition of our $e^{2\gamma_E}$ terms. The result for ${\cal C}(\alpha,z^2\mu^2)$  in \eq{ps-c} should be used to extract an $\overline{\rm MS}$ PDF from an $\overline{\rm MS}$ result for the pseudo-PDF.}
\begin{align} \label{eq:ps-c}
{\cal C}(\alpha,z^2\mu^2)
	=& \left[ 1+ {\alpha_sC_F\over 2\pi}\left({3\over2}\ln(z^2\mu^2)+{3\over2}\ln{e^{2\gamma_E}\over 4}+{3\over2}\right)\right]\delta(1-\alpha)
 \\
 	&+ {\alpha_sC_F\over 2\pi}\left\{\left({1+\alpha^2\over 1-\alpha}\right)^{[0,1]}_{+(1)}\left[-\ln(z^2 \mu^2)-\ln{e^{2\gamma_E}\over 4}-1\right] - \left(4\ln(1-\alpha)\over 1-\alpha\right)^{[0,1]}_{+(1)}  +2(1-\alpha)\right\}\theta(\alpha)\theta(1-\alpha)
  \,. \nn
\end{align}
This result is independent of the infrared regulator as it must be.
We have also computed the matching coefficient for the $\Gamma=\gamma^z$ case, and it is ${\cal C}_{\gamma^z}(\alpha,z^2\mu^2) = {\cal C}(\alpha,z^2\mu^2) + \Delta {\cal C}_{\gamma^z}(\alpha,z^2\mu^2)$ with
\beq
\Delta {\cal C}_{\gamma^z}(\alpha,z^2\mu^2) = {\alpha_sC_F\over 2\pi} 2(1-\alpha)\theta(\alpha)\theta(1-\alpha)\,.
\eeq
Due to the $\ln(z^2\mu^2)\delta(1-\alpha)$ term in \eq{ps-c}, the matching coefficient for the $\overline{\rm MS}$ pseudo-PDF again displays the fact that there is not a smooth local limit as $z\to 0$.  It is possible to define a scheme other than $\overline{\rm MS}$ to ensure that this limit is smooth, reproducing a renormalization for $z\to 0$ that agrees with the fact that the local operator corresponds with a conserved current. One such scheme would be to simply multiply all $\overline{\rm MS}$ renormalization constants by $C_0(\mu^2 z^2)$, which would lead to a   \spcorr renormalized in a different scheme, and a corresponding different matching coefficient in \eq{ps-c} with a smooth $z\to 0$ limit. This is equivalent to studying the ratio of \eq{Qtratio} from the start as advocated in Ref.~\cite{Radyushkin:2017cyf,Radyushkin:2017lvu}. We will give explicit results for this scheme choice below. This modified scheme should not be confused with the strict definition of the $\overline{\rm MS}$ scheme.

From \eq{fac-tq} the corresponding relation for the matching coefficient for the quasi-PDF defined in Eq.~(\ref{eq:fac-tq-orig0}) is
\beq \label{eq:quasiC}
C^{(1)}(\xi,\mu/(|y|P^z))= \tilde q'^{(1)}(\xi,\mu/|y|P^z,\epsilon_{\mbox{\tiny IR}})-q^{(1)}(\xi,\epsilon_{\mbox{\tiny IR}})\,.
\eeq
Therefore using \eq{qPDFft} the matching coefficient relating the quasi-PDF and PDF is
\begin{align} \label{eq:quasi-c}
C\left(\xi, {\mu\over |y| P^z}\right)
 = &\, \delta\left(1-\xi\right)+{\alpha_sC_F\over 2\pi}\left\{
	\begin{array}{ll}
	\displaystyle \left({1+\xi^2\over 1-\xi}\ln {\xi\over \xi-1} + 1 + {3\over 2\xi}\right)^{[1,\infty]}_{+(1)}- \left({3\over 2\xi}\right)^{[1,\infty]}_{+(\infty)}
    &\, \xi>1
    \nn\\[10pt]
	\displaystyle \left({1+\xi^2\over 1-\xi}\left[
     - \ln{\mu^2\over y^2P_z^2} + \ln\big(4\xi(1-\xi)\big)\right] - {\xi(1+\xi)\over 1-\xi}\right)^{[0,1]}_{+(1)}
    &\, 0<\xi<1
    \nn\\[10pt]
	\displaystyle  \left(-{1+\xi^2\over 1-\xi}\ln {-\xi\over 1-\xi} - 1 + {3\over 2(1-\xi)}\right)^{[-\infty,0]}_{+(1)} - \left({3\over 2(1-\xi)}\right)^{[-\infty,0]}_{+(-\infty)} \quad
    &\, \xi<0
	\end{array}\right.\nn\\[5pt]
	& + {\alpha_sC_F\over 2\pi}\left[\delta(1-\xi) \left( {3\over2}\ln{\mu^2\over 4y^2P_z^2} + {5\over2}\right) + {3\over2}\gamma_E\left({1\over(\xi-1)^2}\delta^+({1\over \xi-1}) + {1\over(1-\xi)^2}\delta^+({1\over 1-\xi})\right)\right]\,.
\end{align}
Again this result is independent of the IR regulator as it must be. Here the plus function terms $\big[ g_1(\xi) \big]_{+(1)}^{[1,\infty]}$ and $\big[ g_2(\xi) \big]_{+(1)}^{[-\infty,0]}$ have integrands that converge for $\xi\to \pm \infty$, behaving as $g_i(\xi)\sim 1/\xi^2$.  Note that if we had instead used the renormalized $\overline{\rm MS}$ quasi-PDF calculated in Eq.~(\ref{eq:qPDFren}), we would obtain a different matching coefficient $C$ with different $\delta$-functions at $\xi=\pm\infty$. However, the $\delta$-functions do not contribute to the convolution integral in Eq.~(\ref{eq:fac-tq}) for any integrable PDFs. For example, to carry out the convolution with $1/\xi^2 \delta^+(1/\xi)$ we can use $\delta^+\big({1\over \xi}\big)  = \lim_{\beta\to0^+}  \delta\big({1\over \xi} - \beta\big)$, which
when plugged into the factorization formula gives
\begin{align}
&\lim_{\beta\to0^+}\int {dy\over |y|} {y^2\over x^2}
\, \delta\Big({y\over x} -\beta\Big) f_{u-d}(y)
=\lim_{\beta\to0^+}\beta f_{u-d}(\beta x)\,.
\end{align}
For the plus-function at $\infty$ using \eqs{plusinfinity}{plus1} we have
\begin{align}
\int_{-1}^{+1}\! {dy\over |y|} \left[{1\over (x/y)}\right]_{+(\infty)}^{[1,\infty]} f_{u-d}(y)
&= \lim_{\beta\to 0^+} \int_{-1}^{+1}\! {dy\over |y|}
\bigg[  {\theta(x/y-\beta) \over x/y}
+ {y^2\over x^2}\delta\Bigl({y\over x} -\beta\Bigr)\ln\beta  \bigg] f_{u-d}(y)
\nn\\
&= \int_{-1}^{+1}\! {dy\over x} {y\over |y|} f_{u-d}(y)
+ \lim_{\beta\to 0^+}\beta f_{u-d}(\beta x)\ln\beta
\,. \nn
\end{align}
In the last line we dropped the $\theta(x/y-\beta)$ since at small $y$ our PDF behaves as $f_{u-d}(y)\sim y^{-1+a}$ with $0<a<1$. This also implies
\begin{align}
\lim_{\beta\to0}\beta f_{u-d}(\beta x) \propto\: & x^{-1+a}\lim_{\beta\to0} \beta^a  =0\,, \nn\\
\lim_{\beta\to0}\beta f_{u-d}(\beta x)\ln\beta \propto\: & x^{-1+a}\lim_{\beta\to0} \beta^a \ln\beta =0\,,
\end{align}
which means that the distribution contributions evaluated at $\xi=\pm\infty$ in the matching coefficient $C$ give zero contribution.

Therefore, the matching coefficients calculated from the quasi-PDFs in Eq.~(\ref{eq:qPDFren}) and Eq.~(\ref{eq:qPDFft}) are the same in effect, and we can simply drop all the $\delta$-functions at $\xi=\pm\infty$ when plugging them into the factorization formula:
\begin{align} \label{eq:quasi-matching}
C^{\overline{\rm MS}}\left(\xi, {\mu\over |y| P^z}\right)
= &\, \delta\left(1-\xi\right)+{\alpha_sC_F\over 2\pi}\left\{
\begin{array}{ll}
\displaystyle \left({1+\xi^2\over 1-\xi}\ln {\xi\over \xi-1} + 1 + {3\over 2\xi}\right)^{[1,\infty]}_{+(1)}- {3\over 2\xi}
&\, \xi>1
\nn\\[10pt]
\displaystyle \left({1+\xi^2\over 1-\xi}\left[
- \ln{\mu^2\over y^2P_z^2} + \ln\big(4\xi(1-\xi)\big)\right] - {\xi(1+\xi)\over 1-\xi}\right)^{[0,1]}_{+(1)}
&\, 0<\xi<1
\nn\\[10pt]
\displaystyle  \left(-{1+\xi^2\over 1-\xi}\ln {-\xi\over 1-\xi} - 1 + {3\over 2(1-\xi)}\right)^{[-\infty,0]}_{+(1)} - {3\over 2(1-\xi)}\quad
&\, \xi<0
\end{array}\right.\nn\\[5pt]
& + {\alpha_sC_F\over 2\pi}\delta(1-\xi) \left( {3\over2}\ln{\mu^2\over 4y^2P_z^2} + {5\over2}\right)\,.
\end{align}
The use of \eq{quasi-matching} in the factorization formula is valid for any PDF that behaves as $\lim_{y\to 0} f(y,\mu) \sim y^{-1+a}$ with $a>0$.
We have also computed the matching coefficient for the $\Gamma=\gamma^z$ case, and it is given by $C_{\gamma^z}(\xi, \mu/(|y| P^z)) = C(\xi, \mu/(|y| P^z)) + \Delta C_{\gamma^z}(\xi, \mu/(|y| P^z))$ with
\begin{align}
\Delta C_{\gamma^z}(\xi, \mu/(|y| P^z))  = {\alpha_s C_F\over 2\pi} 2(1-\xi)  
  \,\theta(\xi)\theta(1-\xi) \,.
\end{align}

Note that our result for the quark matching coefficient in $\overline{\rm MS}$ differs from that of Ref.~\cite{Wang:2017qyg} which is a pure plus function, but gives a convolution that does not converge, just as in the case of the quasi-PDF with a transverse momentum cutoff, see Ref.~\cite{Stewart:2017tvs}.

Since the renormalized pseudo-PDF and quasi-PDF satisfy the relation in Eq.~(\ref{eq:equiv}) by definition, $C\left(\xi, \mu/(|y| P^z)\right)$ and ${\cal C}(\alpha,z^2\mu^2)$ that are given by Eqs.~(\ref{eq:ps-c}, \ref{eq:quasi-c}) automatically satisfy the relation in Eq.~(\ref{eq:ftc-C}). 

Besides, if one uses a scheme other than $\overline{\rm MS}$ for the quasi-PDF, such as the scheme obtained by absorbing $C_0$ into the $\overline{\rm MS}$ renormalization constant, then this will lead to a result for the matching coefficient that is a pure plus function and hence satisfies current conservation. Starting with \eq{quasi-c} and using \eq{C0} together with \eq{subtlety} we obtain
\begin{align} \label{eq:quasi-matching-alt}
C^{\rm ratio\!}\left(\xi, {\mu\over |y| P^z}\right)
= &\, \delta\left(1-\xi\right)+{\alpha_sC_F\over 2\pi}\left\{
\begin{array}{ll}
\displaystyle \left({1+\xi^2\over 1-\xi}\ln {\xi\over \xi-1} + 1 - {3\over 2(1-\xi)}\right)^{[1,\infty]}_{+(1)}
&\, \xi>1
\nn\\[10pt]
\displaystyle \left({1+\xi^2\over 1-\xi}\left[
- \ln{\mu^2\over y^2P_z^2} + \ln\big(4\xi(1-\xi)\big)-1\right] +1+ {3\over 2(1-\xi)}\right)^{[0,1]}_{+(1)}
&\, 0<\xi<1
\nn\\[10pt]
\displaystyle  \left(-{1+\xi^2\over 1-\xi}\ln {-\xi\over 1-\xi} - 1 + {3\over 2(1-\xi)}\right)^{[-\infty,0]}_{+(1)}\quad
&\, \xi<0
\end{array}\right.\,,\nn\\
\end{align}
and for the $\Gamma=\gamma^z$ case,
\beq
\Delta C^{\rm ratio}_{\gamma^z}(\xi, \mu/(|y| P^z))  = {\alpha_s C_F\over 2\pi} \big[2(1-\xi)\big]^{[0,1]}_{+(1)} \,.
\eeq
While retaining current conservation in the renormalized quasi-PDF, \eq{quasi-matching-alt} can be used for example as input to the two-step matching procedure in the lattice calculation of PDF in Refs.~\cite{Alexandrou:2017huk}. 
For the matching step, an equivalent procedure is to study the ratio given in \eq{Qtratio} in the $\overline{\rm MS}$ scheme from the start, as advocated in Ref.~\cite{Radyushkin:2017cyf,Radyushkin:2017lvu}, performing its matching onto the PDF, which will yield \eq{quasi-matching-alt}. This concludes our discussion of matching results and the equivalence between the quasi-PDF and pseudo-PDF at one-loop order.
	
\end{widetext}

\section{Other Renormalization Schemes}
\label{sec:ren}

Although we derive the above matching formula assuming that the quasi-PDF is renormalized in the $\overline{\ensuremath{\operatorname{MS}}}$ scheme, this is not a limitation to our result.
Since the gauge-invariant Wilson line operator $\tilde{O}_\Gamma(z)$ has been proven to be multiplicatively renormalizable in the coordinate space~\cite{Ji:2017oey,Ishikawa:2017faj}, one can convert $\tilde{Q}_{\Gamma}(z)$ from any other scheme to the $\overline{\ensuremath{\operatorname{MS}}}$ scheme before using the above factorization formula. The renormalization of the
quasi-PDF has been studied in many recent papers~\cite{Ji:2015jwa,Ishikawa:2016znu,Chen:2016fxx,Xiong:2017jtn,Constantinou:2017sej,Alexandrou:2017huk,Chen:2017mzz,Green:2017xeu,Stewart:2017tvs,Wang:2017eel}. We will discuss some
of these results and show how they can be incorporated into the factorization formula in Eq.~(\ref{eq:fac-tq}).

The $\overline{\ensuremath{\operatorname{MS}}}$ scheme is convenient for our discussion of the OPE as it guarantees Lorentz and gauge invariances, but it is not practical for lattice renormalization. Since the lattice theory has a natural UV cut-off $1/a$ with $a$ being the lattice spacing, the unrenormalized \spcorr $\tilde{Q}$ inherits the power divergence from the Wilson line self-energy according to Eq.~(\ref{eq:ren}). For an arbitrary scheme $X$, the renormalized \spcorr
\begin{align}
\tilde{Q}^X(\zeta,z^2\mu^2_R)
  = \lim_{a\to0} Z^{-1}_X(z^2\mu_R^2,a^2\mu^2_R)\, \tilde{Q}(\zeta,z^2/a^2)
\end{align}
should be free of all the UV divergences and have a well-defined continuum limit as $a\to0$. This continuum limit, in particular, is independent of the UV regulator, so
\begin{align}
&\lim_{a\to0} Z^{-1}_X(z^2\mu_R^2,a^2\mu^2_R)\tilde{Q}(\zeta,z^2/a^2) \nn\\
&= Z^{-1}_X(z^2\mu^2_R,\epsilon)\tilde{Q}(\zeta,z^2,\epsilon)\,.
\end{align}
As a result, we can relate $\tilde{Q}^X(\zeta,z^2\mu^2_R)$ to the $\overline{\rm MS}$ scheme by the conversion
\begin{align}
\tilde{Q}^X(\zeta,z^2\mu^2_R) =& \frac{Z_{\overline{\rm MS}}(\epsilon,\mu)}{Z_X(z^2\mu^2_R,\epsilon)}\, \tilde{Q}^{\overline{\rm MS}}(\zeta,z^2\mu^2)\nn\\
=& Z'_X(z^2\mu_R^2,\mu_R^2/\mu^2)\, \tilde{Q}^{\overline{\rm MS}}(\zeta,z^2\mu^2)\,,
\end{align}
where the regulator $\epsilon$ dependence is completely canceled out between $Z_{\overline{\rm MS}}$ and $Z_X$. The ratio $Z_X'$ can be calculated perturbatively in QCD, which was done in~\cite{Constantinou:2017sej} for several lattice schemes and the RI/MOM scheme. Thus the factorization formula we have proven in Sec.~\ref{sec:ope} still applies to $\tilde{Q}^X$ with a slight modification to the coefficient function,
\begin{align}
\tilde{Q}^X(\zeta,z^2\mu^2_R)
 = \int_{-1}^1 \!\! d\alpha \:
   \mathcal{C}^X(\alpha,\mu_R^2/\mu^2,\mu^2 z^2)\, Q(\alpha\zeta,\mu)\,,
\end{align}
where the matching coefficient for the scheme $X$ is related to that of $\overline{\rm MS}$ by
\begin{align}
\mathcal{C}^X(\alpha,\mu_R^2/\mu^2,\mu^2 z^2) = Z'_X(z^2\mu_R^2,\mu_R^2/\mu^2)\, \mathcal{C}(\alpha,\mu^2 z^2) \,.
\end{align}
For the pseudo-PDF the modified result also involves this same coefficient
\begin{align}
\mathcal{P}^X(x,z^2\mu^2_R)
=&\int_{|x|}^1 {dy\over |y|}\
 \mathcal{C}^X\Bigl({x\over y},{\mu_R^2\over \mu^2},\mu^2 z^2\Bigr) q(y,\mu)
  \\
&+\int^{-|x|}_{-1} {dy\over |y|}\
 \mathcal{C}^X\Bigl({x\over y},{\mu_R^2\over \mu^2},\mu^2 z^2\Bigr) q(y,\mu)
 \,. \nn
\end{align}
Meanwhile, for the quasi-PDF we have,
\begin{align}\label{eq:rimomfac}
\tilde{q}_X\left(x,{\mu_R^2\over P_z^2}\right)
&\equiv \int \frac{d \zeta}{2 \pi}\: e^{ix \zeta}\:  \tilde{Q}^X \biggl(\zeta, \frac{{\mu}^2\zeta^2}{P_z^2}\biggr)
\nn\\
&=\int_{-1}^1 {dy\over |y|}
\: C^X\Bigl(\frac{x}{y},\frac{\mu_R}{\mu},\frac{\mu}{|y| P^z}\Bigr)\, q(y,\mu)\,.
\end{align}
Here the modified coefficient for the $X$ scheme is related to coefficient in the $\overline{\rm MS}$ scheme by
\begin{align}
 & C^X\Bigl(\frac{x}{y},  \frac{\mu_R}{\mu},\frac{\mu}{|y| P^z}\Bigr)
 \\
&\ =\!\! \int\!\! d\eta\ \bar{Z}'_X\Bigl(\eta^2, {\mu_R^2\over \mu^2}\Bigr)\:
 C\Bigl( \frac{x}{y} - \frac{\eta}{|y|}\frac{\mu_R}{P^z},{\mu\over |y|P^z}\Bigr)
 , \nn
\end{align}
where here $\bar{Z}'_X$ is defined by the Fourier transform
\begin{align}
\bar{Z}'_X\left( \eta^2,{\mu_R^2\over \mu^2}\right)
\equiv  \int {d\tau\over 2\pi}\ e^{i\eta \tau }\, Z'_X\Bigl(\tau^2,\frac{\mu_R^2}{\mu^2}\Bigr)
  \,.
\end{align}
Depending on the scheme $X$ we note that slightly modified definitions of $\bar{Z}'_X$ may be more appropriate.

One undesirable feature of  the $\overline{\rm MS}$ scheme for the renormalized \spcorr is that it does not have a smooth $z\to 0$ limit, and hence no simple connection with the fact that the local operator for $z=0$ is a conserved current. To avoid this one can simply make use of a different scheme that has a simple relation to $\overline{\rm MS}$, such as by adding $C_0(\mu^2z^2)$ to the $\overline{\rm MS}$ renormalization constant. This removes the offending $\ln(\mu^2 z^2)$ terms and yields a scheme with a smooth connection to the conserved current.

\begin{figure*}
	\centering
	\includegraphics[width=.49\textwidth]{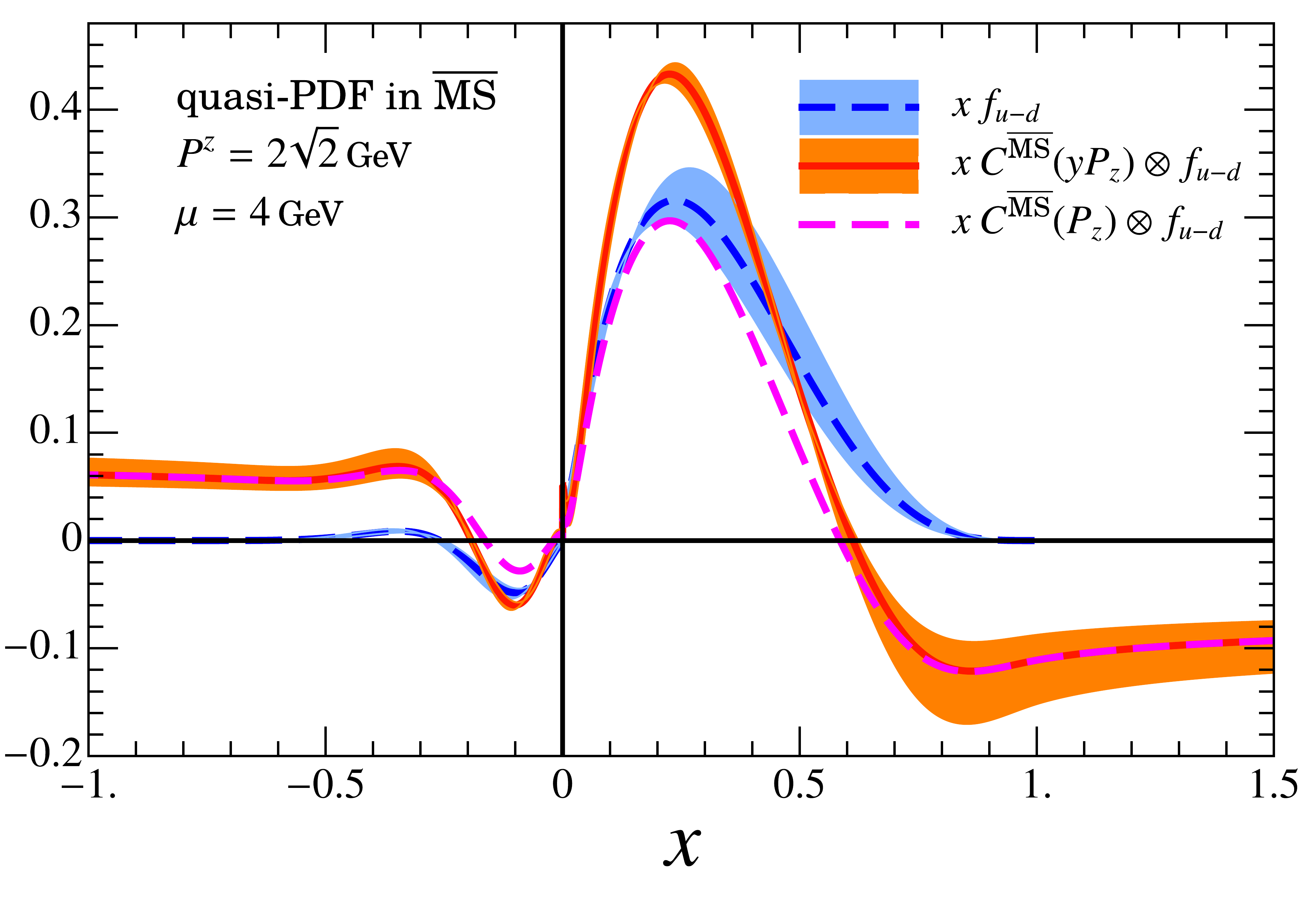}
	\vspace{-0.3cm}	
	\caption{
		The $\overline{\rm MS}$ scheme PDF $xf_{u-d}$ and the $\overline{\rm MS}$ quasi-PDF obtained from $x\, C^{\overline{\rm MS}}(p^z)\otimes f_{u-d}$, comparing results obtained with $p^z=yP^z$ and $p^z=P^z$.
	}
	\label{fig:quasi}
\end{figure*}

\begin{figure*}
	\centering
	\includegraphics[width=.49\textwidth]{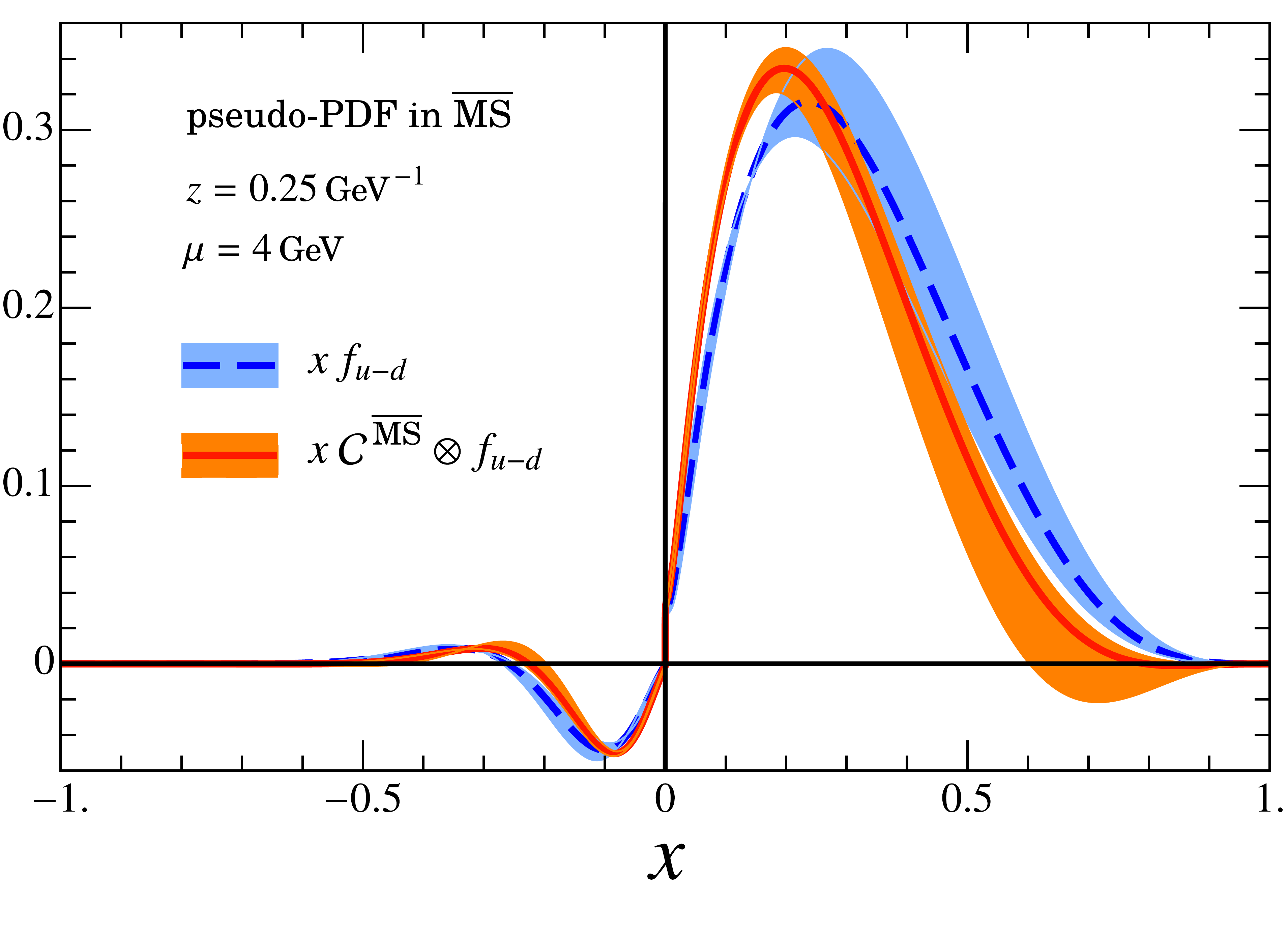}
	\hspace{0.1cm}
	\includegraphics[width=0.49\textwidth]{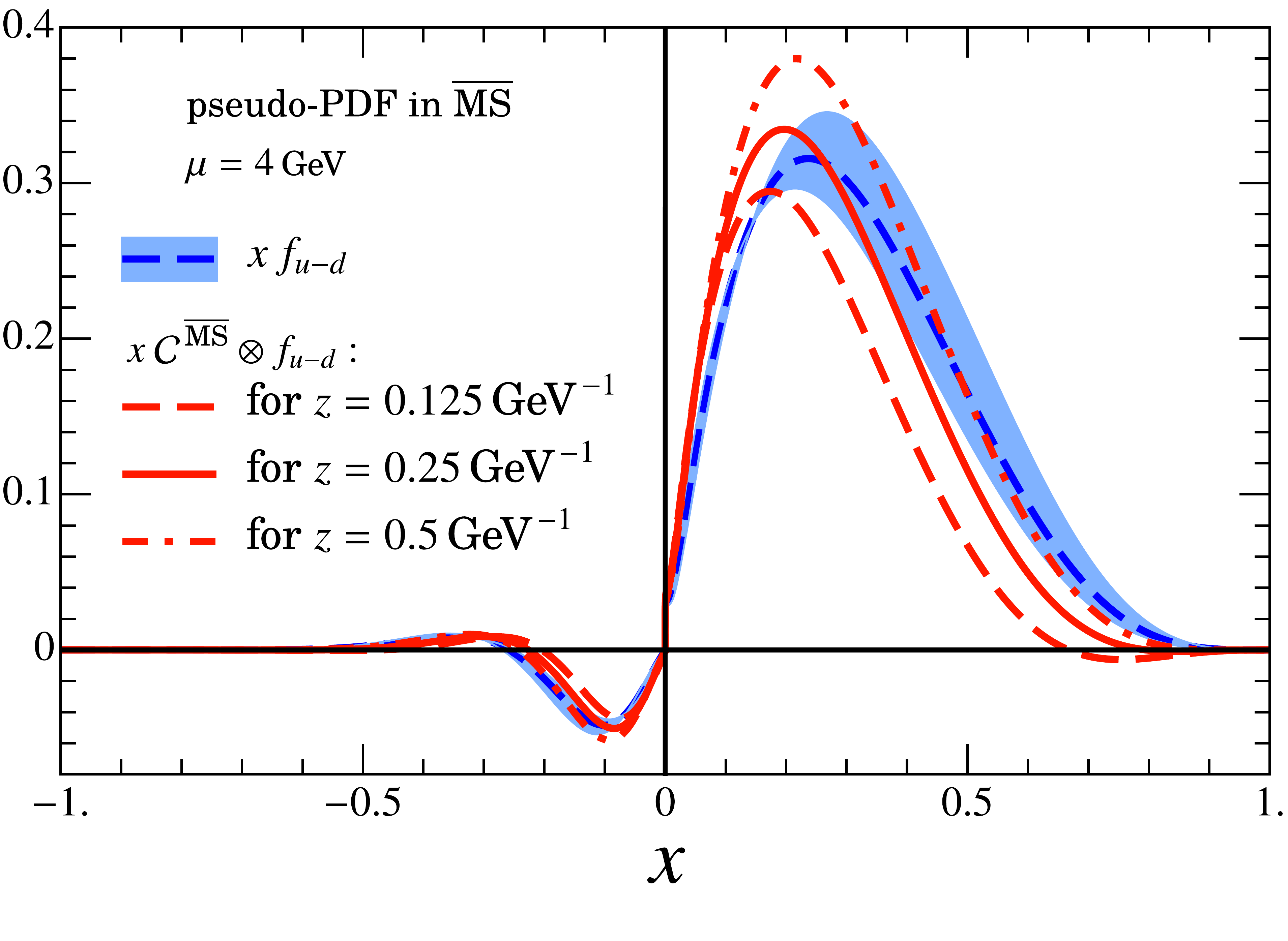}	\\
	\vspace{-0.3cm}
	\caption{(Left) Comparison between the  PDF $xf_{u-d}$ and the pseudo-PDF  $x({\cal C}^{\overline{\rm MS}}\otimes f_{u-d})$ in the $\overline{\rm MS}$ scheme. The orange and blue bands indicate the results from varying the factorization scale $\mu=4\,{\rm GeV}$ by a factor of two. (Right) Same but now showing only central pseudo-PDF curves for different values of $z$. }
	\label{fig:pseudo}
\end{figure*}

Besides the $\overline{\rm MS}$ scheme, the quasi-PDF has also been defined with a transverse momentum cut-off~\cite{Xiong:2013bka,Lin:2014zya,Ma:2014jla,Alexandrou:2015rja} and in the RI/MOM scheme~\cite{Constantinou:2017sej,Alexandrou:2017huk,Chen:2017mzz,Green:2017xeu,Stewart:2017tvs}.  The RI/MOM scheme has attracted strong interest recently as it can be implemented nonperturbatively on the lattice, so we consider it as an explicit example of the above relations. In this scheme, the renormalization constant $Z_{\rm OM}$ is determined by imposing a condition on the \spcorr in an off-shell quark state,
\begin{align}
&\left.Z^{-1}_{\rm OM}\,
\tilde{Q}(\zeta=zp^z,z^2/a^2,-p^2a^2)\right|_{p^2=-\mu_R^2,p^z=p_R^z}
\nn\\
& \quad = \tilde{Q}^{(0)}_q(zp_R^z,z^2/a^2,z^2\mu_R^2) = e^{-izp_R^z}\,,
\end{align}
where $q$ denotes the quark state, $p^\mu$ is the external momentum, and ``$(0)$" in the superscript stands for the tree-level matrix element. As a result,
\begin{align}
Z_{\rm OM} &= Z_{\rm OM} (zp_R^z, z^2/a^2,a^2\mu_R^2)\,,\nn\\
Z'_{\rm OM}&=Z'_{\rm OM}(zp_R^z, z^2\mu_R^2, \mu_R^2/\mu^2)\,,
\end{align}
and here we define
\begin{align}
	&\bar{Z}'_{\rm OM}\left(\eta,{\mu_R^2\over (p_R^z)^2},{\mu_R^2\over \mu^2}\right)\nn\\
	&\equiv
   p_R^z \int {dz\over 2\pi}\ e^{i\eta p_R^z z}\ Z'_{\rm OM}(zp_R^z, z^2\mu_R^2, \mu_R^2/\mu^2)\,.
\end{align}
Then the matching coefficient in Eq.~(\ref{eq:rimomfac}) becomes
\begin{align}
& C^{\rm OM}\left(\frac{x}{y}, {\mu_R\over p_R^z},{\mu_R\over \mu}, {\mu\over yP^z}\right)
 \\
=& \int d\eta\ \bar{Z}'_{\rm OM}\left(\eta,{\mu_R^2\over (p_R^z)^2},{\mu_R^2\over\mu^2}\right) C\left({x\over y}-{\eta\over y}{p_R^z\over P^z}, {\mu\over |y|P^z}\right)
 \,. \nn
\end{align}
The choices of $\mu_R$ and $p_R^z$ are independent of $\mu$ and $P^z$, and $p_R^z=P^z$ was used in Refs.~\cite{Chen:2017mzz,Stewart:2017tvs}.

It should be noted that on the lattice, due to the breaking of chiral symmetry, the vector-like quark Wilson line operator $\tilde{O}_{\gamma^\mu}(z)$ can mix with the scalar operator $\tilde{O}_{\bf 1}(z)$, as has been discussed in Refs.~\cite{Constantinou:2017sej,Alexandrou:2017huk,Chen:2017mzz,Green:2017xeu,Chen:2017mie}. After considering the mixing effects, the same factorization formula can still be applied to the RI/MOM quasi-PDF from lattice QCD.

\section{Numerical results}
\label{sec:num}

In this section we numerically analyze the quasi-PDF, \spcorr and pseudo-PDF by studying how the matching coefficients in Eqs.~(\ref{eq:fac-tq},\ref{eq:ps-q-fact}) change the PDF. The quasi-PDF has already been studied in this manner for the $\overline{\rm MS}$, transverse momentum cut-off, and RI/MOM schemes in Ref.~\cite{Stewart:2017tvs}. Our  new $\overline{\rm MS}$ result for the matching is given in \eq{quasi-matching}, and leads to stable convolution integrals.  We also compare the differences between using hadron momentum $p^z=P^z$ and the parton momentum $p^z=|y|P^z$ for the matching coefficient in the $\overline{\rm MS}$ scheme. We take $\Gamma =\gamma^0$ for the results here.

As an example we use for our analysis the unpolarized iso-vector parton distribution,
\begin{align}
f_{u-d}(x,\mu) =  f_u(x,\mu) - f_d(x,\mu) - f_{\bar{u}}(-x,\mu) +  f_{\bar{d}}(-x,\mu),
\end{align}
where we include
$f_{\bar{u}}(-x,\mu) = - f_{\bar{u}}(x,\mu)$ and $f_{\bar{d}}(-x,\mu) = - f_{\bar{d}}(x,\mu)$, the anti-parton distributions. For ease of comparison, we use the next-to-leading-order iso-vector PDF $f_{u-d}$ from MSTW 2008~\cite{Martin:2009iq} with the corresponding running coupling $\alpha_s(\mu)$.

To implement the plus functions in the numerical calculation, we impose a soft cutoff $|y-x|<10^{-m}$ and test the sensitivity of results to $m$. Since the limit of $y\to0$ corresponds to the asymptotic region $|x/y|\to\infty$, we also impose a UV cutoff $|y|>10^{-n}$ to test the convergence of the convolution integral. We find that all the results presented below are insensitive to $m$ and $n$. The fact that our result in \eq{quasi-matching} has terms outside the plus function at $1$ in each of the $\xi\in [1,\infty]$ and $\xi\in [-\infty,0]$ intervals is important for ensuring that our $\overline{\rm MS}$ result for $C$ is insensitive to the $|y|>10^{-n}$ cutoff. This was not the case for the quasi-PDF that was defined with a transverse momentum cutoff~\cite{Xiong:2013bka}. The RI/MOM scheme result~\cite{Stewart:2017tvs} also does not suffer from this issue.

In Fig.~\ref{fig:quasi} we compare the PDF with the quasi-PDF in the $\overline{\rm MS}$ scheme obtained from the convolution in \eq{fac-tq} using our one-loop result in \eq{quasi-matching}. We observe that changing from $p^z=P^z$  to the correct $p^z=|y|P^z$ shifts the result in the physical region by a considerable amount.

\begin{figure*}
	\centering
	\includegraphics[width=.49\textwidth]{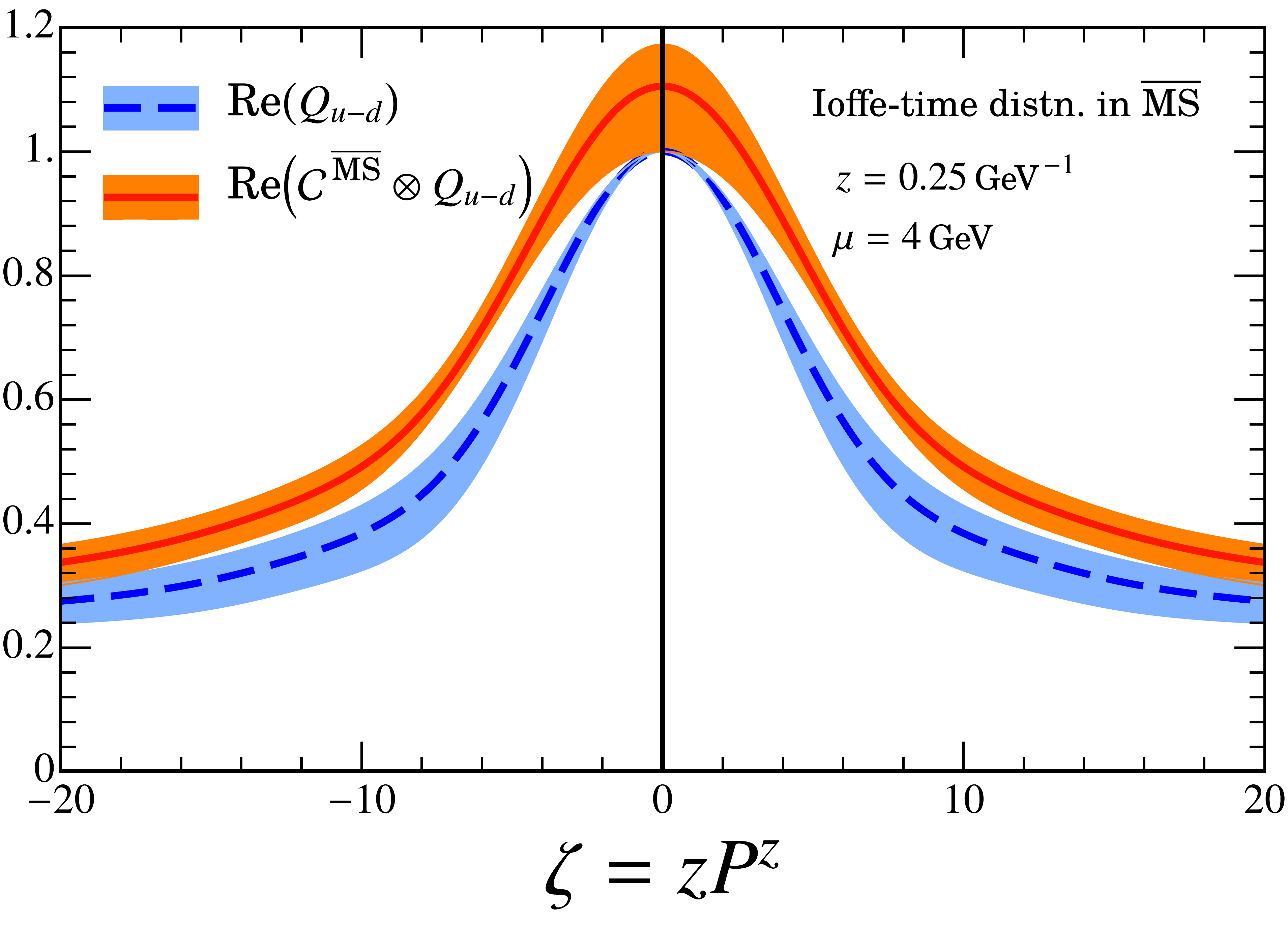}
	\hspace{0.1cm}
	\includegraphics[width=0.49\textwidth]{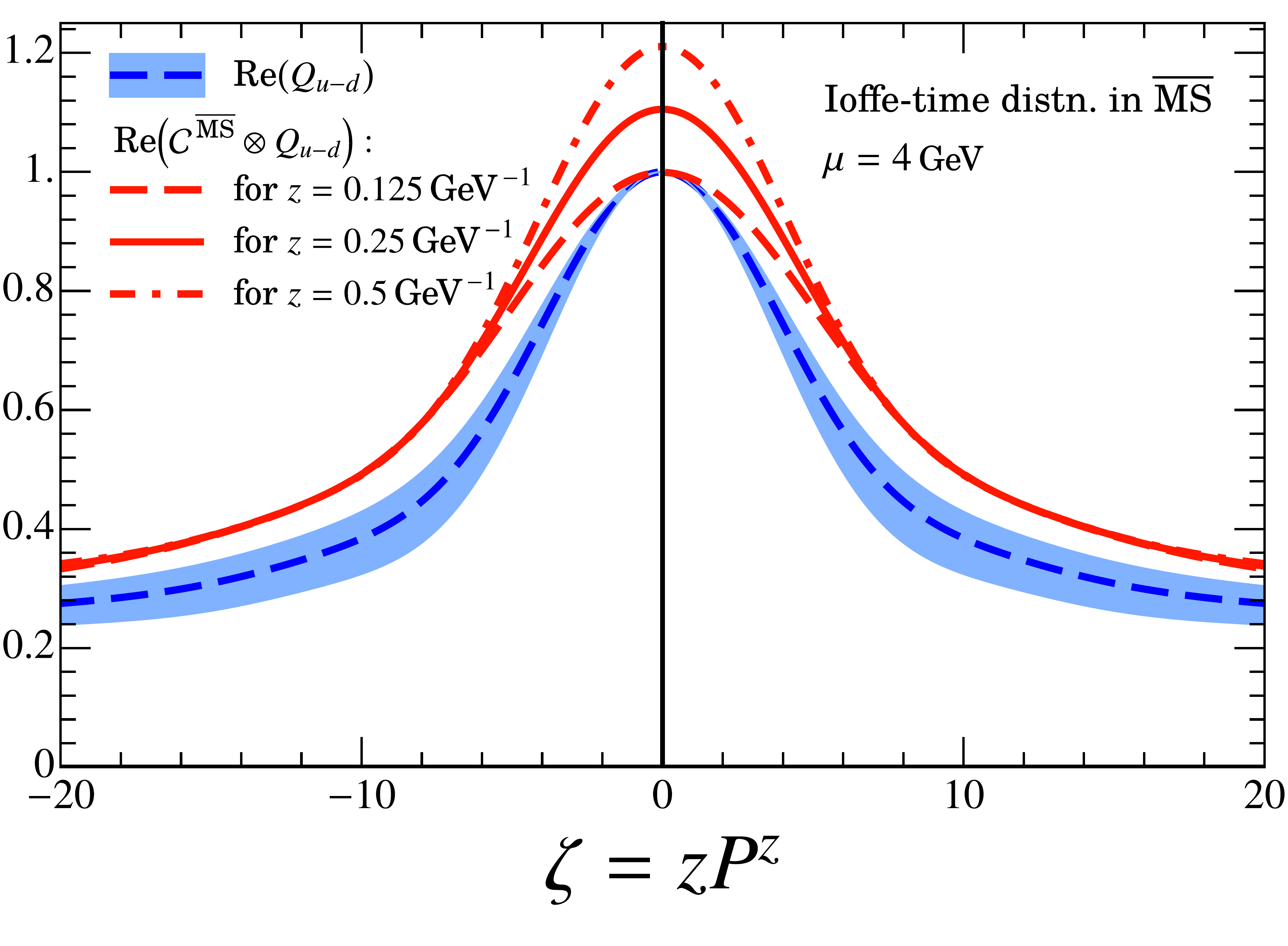}
	\\
	\includegraphics[width=.49\textwidth]{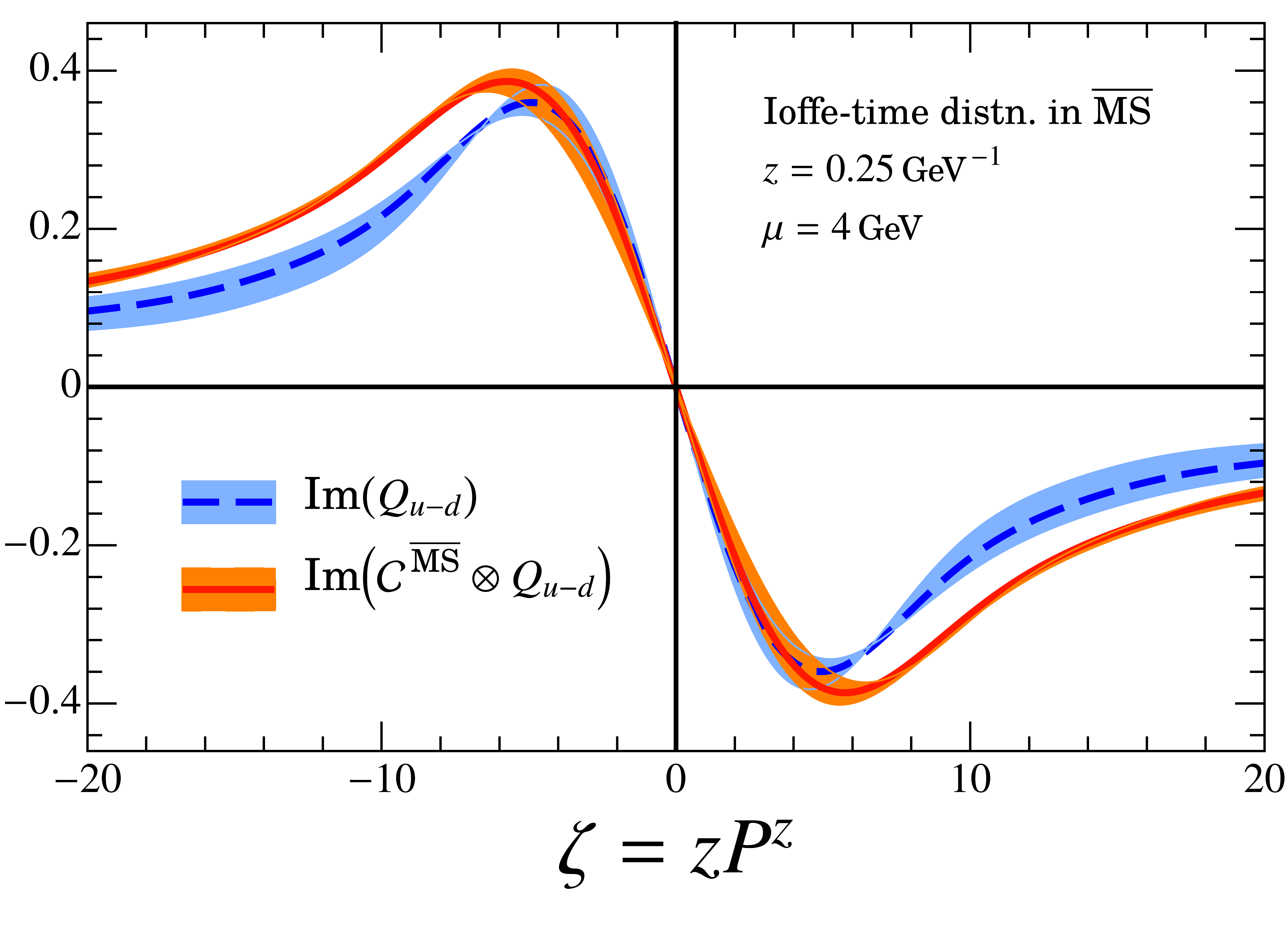}
	\hspace{0.1cm}
	\includegraphics[width=0.49\textwidth]{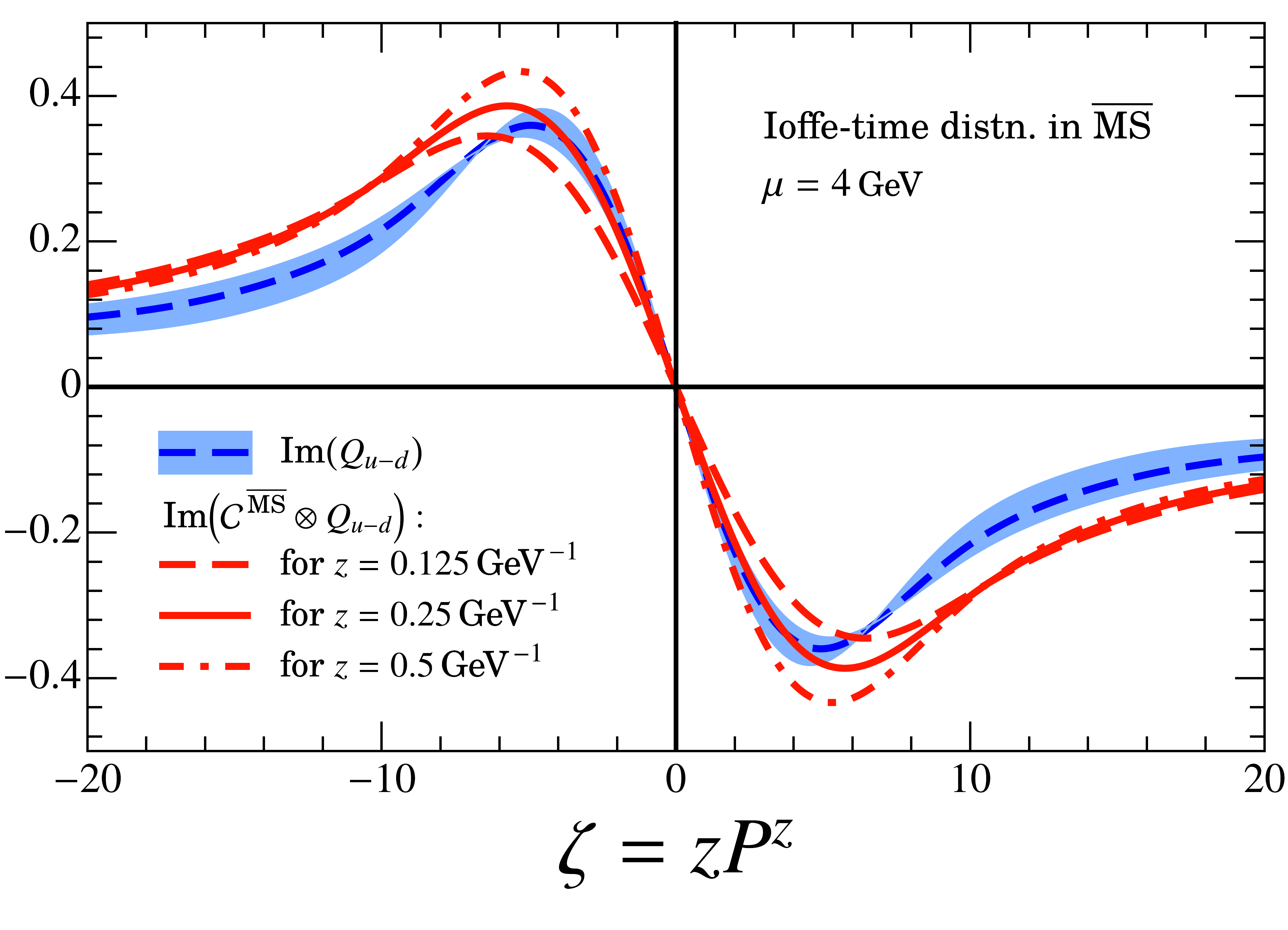}	\\
\vspace{-0.3cm}
	\caption{(left) Comparison between the light-cone time distribution $Q_{u-d}$ and \spcorr from $({\cal C}^{\overline{\rm MS}}\otimes Q_{u-d})$ in the $\overline{\rm MS}$ scheme. The orange and blue bands indicate the results from varying the factorization scale $\mu=4\,{\rm GeV}$ by a factor of two. (right) Same but now showing only central \spcorr curves for different values of $z$. The top panels show the real part, while bottom panels show the imaginary part. }
	\label{fig:ioffe}
\end{figure*}

The same type of comparison can be made for the pseudo-PDF in the $\overline{\rm MS}$ scheme by applying the factorization formula in Eq.~(\ref{eq:ps-q-fact}) and matching coefficient in Eq.~(\ref{eq:ps-c}). In Fig.~\ref{fig:pseudo}a we compare the PDF and pseudo-PDF and their dependence on the factorization scale $\mu$, while in Fig.~\ref{fig:pseudo}b we include the dependence of the pseudo-PDF on the distance $|z|$. Since the matching coefficient in Eq.(\ref{eq:ps-c}) is similar to the parton splitting function except for the nontrivial finite constants, matching the PDF to the pseudo-PDF is analogous to evolving the PDF from $\mu$ to the scale of $1/|z|$. This evolution has been calculated in Refs.~\cite{Radyushkin:2017cyf}.  The variation of $|z|$ has a similar effect to the PDF evolution, as is observed in the right panel of Fig.~\ref{fig:pseudo}. When $|z|\mu=1$, the logarithm is zero, and the matching effect from ${\cal C}$ is determined by the nontrivial constants in Eq.(\ref{eq:ps-c}), which shifts the PDF downward in the large-$x$ region and upward in the small-$x$ region.

Finally, we can make a similar comparison for the \spcorr in the $\overline{\rm MS}$ scheme obtained with Eq.~(\ref{eq:io-Q-fact}) and Eq.~(\ref{eq:ps-c}). Its real and imaginary parts are even and odd functions of $\zeta$ respectively, and are shown in Fig.~\ref{fig:ioffe}. Again we show the residual dependence on $\mu$ and $|z|$ which are similar to that for the pseudo-PDF.   The matching broadens the curves in the coordinate space.  The \spcorr renormalized in the $\overline{\rm MS}$ scheme does not exhibit vector current (or particle number) conservation, which can be clearly seen from the fact that the real part of the distribution is not equal to 1 at $\zeta=0$ (except for the special case where $|z|\mu$ is tuned to cancel the constant terms in the one-loop ${\cal C}$).

\section{Implications for lattice calculations}
\label{sec:lattice}

Our proof in \sec{ope} makes clear the relationship between the renormalized quasi-PDF, \spcorr, and pseduo-PDF distributions.  As a practical matter there are a few different ways in which these equations can be used to convert a lattice calculation of the \spcorr $\tilde Q$ into a PDF. Three examples are 1) first Fourier transform to the quasi-PDF with \eq{fac-tq-orig0}, and then use \eq{fac-tq}, 2) first Fourier transform to the pseudo-PDF with \eq{pseudoRen}, and then use \eq{ps-q-fact}, and 3) first match to the Fourier transform the position space PDF $Q(\zeta,\mu)$ using \eq{io-Q-fact}, and then transform it to the PDF with the inverse of \eq{FTq}.  Since the numerical implementation of these steps may have slightly different systematics it is interesting to compare them, or to use more than one approach in order to reduce uncertainties.

According to the analysis in Sec.~\ref{sec:ope}, for the factorization formula of the Euclidean distributions to work, one must calculate the same \spcorr with small distance $z^2$ and large momentum $P^z$ so that the dynamical and kinematic higher-twist effects are suppressed. For practical lattice calculations, this means that there is only a finite number of useful data points in $(z, P^z)$ that we can use to extract the PDF.

To illustrate this, consider a $48^3\times64$ lattice with spacing $a=0.09$ fm. The distance of the spatial correlation $z$ is in units of $a\sim1/2.2\ {\rm GeV}^{-1}$, and the nucleon momentum $P^z$ is in units of $2\pi/(48a)\sim0.29$ GeV. Let us take $\Lambda_{\rm QCD}\sim 0.3$ GeV. In principal the target mass corrections can be subtracted. If we consider various values $z=ma$ and $P^z=n*2\pi/(48a)$ for integer $m$ and $n$, then to satisfy $z\Lambda_{\rm QCD}\ll1$ and $P^z\gg \Lambda_{\rm QCD}$, we must have
\beq
 m\ll 11\,,\,\,\,\quad n\gg 1\,.
\eeq
To control the higher-twist correction at $20\%$, i.e. $z^2\Lambda_{\rm QCD}^2\sim0.2,\ \Lambda_{\rm QCD}^2/P_z^2\sim0.2$, we can only choose
\beq
m = \{0,1,2,3,4\}\,,\,\,\,\, n=\{3,4,5,6,\cdots\}\,,
\eeq
where the largest value for $n$ is limited by what is practical in current lattice simulations. Six is the largest number of units attained in Ref.~\cite{Orginos:2017kos}. For quasi-PDF calculations, there are $4\times2+1=9$ useful data points for each fixed momentum $|P^z|$; for pseudo-PDF calculation, there are only $4\times2=8$ useful data points for each fixed $|z|$. In either case, it is anticipated that a direct Fourier transform with respect to $z$ or $\zeta=zP^z$ will lead to oscillation in $x$-space and incorrect prediction for the small-$x$ region due to the truncation in coordinate space. This has been observed in a recent lattice calculation of the quasi-PDF in Ref.~\cite{Chen:2017mzz}. Methods have been developed in recent works to eliminate the oscillation from the truncation effect~\cite{Lin:2017ani,Zhang:2017gau} in the quasi-PDF, while the higher-twist contributions at large $z$ still need to be systematically corrected. It should be noted that the above is a rough estimate of the higher-twist corrections since the prefactor of $z^2\Lambda_{\rm QCD}^2$ could be smaller than 1. Their actual significance can only be quantitatively determined from lattice simulations.

To fully take advantage of all the useful data points, we can evolve them to either the same $z^2$ or $P^z$ according to the perturbative analysis, which has been studied in Refs.~\cite{Orginos:2017kos,Karpie:2017bzm,Radyushkin:2017lvu} for the \spcorr. However, since the evolution equation of the \spcorr in $\ln z^2$ or $\ln P_z^2$ follows a nonlocal convolution in $\zeta=zP^z$ or $z$, one has to know the full information in coordinate space to do the evolution. With limited number of data points, either large uncertainties or adopting a model-dependent assumption about the shape is inevitable.

To improve the precision of either approach, the only way forward is to have finer lattice spacing $a$ so that we could have more data points which satisfy $|z|\ll \Lambda_{\rm QCD}^{-1}$ and larger nucleon momentum $P^z$.
With increasing $P^z$, the valence distribution of the nucleon is contracted in the $z$ direction, so the spatial correlation of valence quarks is shrinked into smaller distance in $z$. If $P^z$ is large enough, the spatial correlation will fall off quickly within $|z|< \Lambda_{\rm QCD}^{-1}$, then the truncation error from Fourier transform will be significantly reduced. On the other hand, if we interpret the spatial correlation as the \spcorr, its shape will not change under a Lorentz boost because it is a scalar function of $\zeta=zP^z$ and $z^2$. Nevertheless, finer lattice spacing $a$ allows for calculation with a wider range of $P^z$, thus covering larger values of $\zeta=zP^z$ to reduce the truncation error. Since the number of useful data points increases quadratically with $1/a$, a more precise lattice calculation with controlled systematic errors will be available in the future.

\section{Conclusions}
\label{sec:summary}

Starting with a Euclidean operator product expansion for products of gauge invariant operators in QCD, we have derived the factorization formulas for the quasi-PDF, \spcorr and pseudo-PDF. The three Euclidean distribution functions are related observables, and all follow from the same fundamental factorization.  For the \spcorr this derivation implies that the ratio in Eq.~(\ref{eq:ratio}) does scale in $z^2$, but needs the small $z^2$ factorization formula in \eq{io-Q-fact} to extract the PDF.
Our derivation for the factorization formula applies when the renormalized \spcorr is defined in any scheme.
The OPE used here could also be used to systematically derive factorization formulas for power corrections to \eq{momfact}, which will involve matching onto higher-twist parton distributions. (The numerical relevance of these corrections is considered in Ref.~\cite{Chen:2016utp}.)
Note that LaMET is not equivalent to the expansion from the OPE, as the former is more general and can be applied to the lattice calculations of other quantities, for example the TMD-PDF where a simple OPE does not exist.

Our derivation of the factorization formula for the quasi-PDF also verifies that the parton momentum $p^z$ in the matching coefficient in Eq.~(\ref{eq:momfact}) has to be $p^z=yP^z$, which makes a considerable difference for the $\overline{\rm MS}$  matching result when compared with $p^z=P^z$ (see Fig.~\ref{fig:quasi}). The proper $p^z$ should therefore be used in lattice calculations of the PDF in the LaMET approach.

As a non-trivial test of relations between the various distributions and factorization formulas we have considered results at one-loop in the $\overline{\rm MS}$ scheme. We have derived a corrected results for the coefficient $C$ for the $\overline{\rm MS}$ scheme, given in \eq{quasi-matching}, which leads to convergent results in the convolution integral.   We have also computed the one-loop $\overline{\rm MS}$ result for the Wilson coefficient ${\cal C}$ appearing in the \spcorr and pseudo-PDF factorizations. A numerical analysis of these one-loop corrections in $\overline{\rm MS}$ has also been provided. The one-loop matching coefficient ${\cal C}$ has a smaller effect for the pseudo-PDF than $C$ does for quasi-PDF, as can seen by comparing \figs{quasi}{pseudo}. Given systematic uncertainties in manipulating the lattice data, it is potentially interesting to consider using the same lattice data on the \spcorr to extract the parton distribution function using both the quasi-PDF and pseudo-PDF approaches.

There are several different ways of implementing the factorization formula to calculate the PDF from lattice data for the \spcorr $\tilde Q$, which we have discussed in \sec{lattice}. One always has to work with short distance correlation and large nucleon momentum to reduce higher-twist corrections. This limits the number of useful data points from lattice calculations as described in \sec{lattice}.  To achieve precision calculations without making model assumptions it will be highly desirable to move towards finer lattice spacing to increase the number of effective data points.

\section*{Acknowledgments}
The authors are thankful for discussions with J.~H.~Zhang and Y.~B.~Yang. This material was partially supported by the U.S. Department of Energy, Office of Science, Office of Nuclear Physics from DE-FG02-93ER-40762, DE-SC0011090 and DE-SC0012704, by the Laboratory Directed Research and Development (LDRD) funding of BNL under contract DE-EC0012704, by a grant from the National Science Foundation of China (No.~11405104), and within the framework of the TMD Topical Collaboration.
I.S. was also supported in part by the Simons Foundation through the Investigator grant 327942.
T.I. was also supported by JSPS KAKENHI Grant Numbers JP26400261, JP17H02906, and also MEXT as ”Priority Issue on Post-K computer” (Elucidation of the Fundamental Laws and Evolution of the Universe) and JICFuS.

\appendix

\section{Fourier Transform}
\label{sec:ft}

To Fourier transform the bare pseudo-PDF in Eq.~(\ref{eq:1looppseudo}) into the bare quasi-PDF, we use the identity
\begin{align} \label{eq:ft}
&\int {d \zeta \over 2\pi}\, e^{i x \zeta} (z^2\mu^2)^{\epsilon} \Gamma(-\epsilon) 4^{-\epsilon}\nn\\
= & \left({\mu^2\over p_z^2}\right)^\epsilon \int {d \zeta \over 2\pi}\, e^{i x \zeta} \int_0^\infty d\alpha\, \alpha^{-1-\epsilon}e^{-\alpha\zeta^2}4^{-\epsilon}\nn\\
=& \left({\mu^2\over p_z^2}\right)^\epsilon \int_0^\infty {d\alpha\over 2\sqrt{\pi}}\, \alpha^{-3/2-\epsilon}e^{-x^2/(4\alpha)}4^{-\epsilon}\nn\\
=& \left({\mu^2\over p_z^2}\right)^\epsilon {\Gamma(\epsilon+1/2)\over \sqrt{\pi}} {1\over |x|^{1+2\epsilon}}\,,
\end{align}
which is true for $\epsilon<0$.

Now let us turn to a plus function $g(y)^{[0,1]}_{+(1)}$. The double Fourier transform
\begin{align}
& \int {d \zeta \over 2\pi}\, e^{i x \zeta} \int_0^1 dy\, e^{-iy\zeta} g(y)^{[0,1]}_{+(1)} (z^2\mu^2)^{\epsilon} \Gamma(-\epsilon) 4^{-\epsilon}\nn\\
= &(i\zeta)\int {d \zeta \over 2\pi}\, e^{i (x-1) \zeta} \int_0^1 dy \int_0^1 dt\ y\,g(1-y)\nn\\
&\times e^{ity\zeta} (z^2\mu^2)^{\epsilon} \Gamma(-\epsilon) 4^{-\epsilon} \nn\\
=& \left({\mu^2\over p_z^2}\right)^\epsilon {\Gamma(\epsilon+{1\over2})\over \sqrt{\pi}}{\partial\over \partial x} \int_0^1 dy \int_0^1 dt \, {y\,g(1-y) \over |x-1+ty|^{1+2\epsilon}}\,.
\end{align}
Since the integration over $y$ and $t$ leads to a piecewise function of $x$, the derivative will end up with a plus function as the discontinuity at $x=1$ gives $\delta(1-x)$.

\section{Quasi-PDF Calculation}
\label{sec:quasi}

Here we quote the pure dimensional regularization results obtained when we carry out the quasi-PDF calculation following Refs.~\cite{Xiong:2013bka,Stewart:2017tvs} by Fourier transforming from $z$ to $x p^z$ for the integrand to obtain $[\delta(p^z-xp^z)-\delta(k^z-xp^z)]$.
For the vertex and wavefunction renormalization graphs we obtain
\begin{widetext}
	
\begin{align}
	&\tilde{q}^{(1)}_\text{vertex}(x,p^z,\epsilon)
	+ \tilde{q}^{(1)}_\text{w.fn.}(x,p^z,\epsilon)\nn\\
	&={\alpha_sC_F\over 2\pi}\left({\mu^2\over p_z^2}\right)^\epsilon \left[ (1-\epsilon)\int_0^1 dy (1-y){1\over |x-y|^{1+2\epsilon}}{\Gamma(\epsilon+{1\over2})\over\sqrt{\pi}}- \delta(1-x){1\over2} \left({1\over\epsilon_{\mbox{\tiny UV}}} - {1\over \epsilon_{\mbox{\tiny IR}}}\right)\right]\,,
\end{align}
while for the sail diagram
\begin{align}
	&\tilde{q}^{(1)}_\text{sail}(x,p^z,\epsilon) \nn\\
	&= {\alpha_sC_F\over 2\pi}\left({4\pi\mu^2\over p_z^2}\right)^\epsilon\left[\int_0^1 dy {x+y\over 1-x}{1\over |x-y|^{1+2\epsilon}}{\Gamma(\epsilon+{1\over2})\over\sqrt{\pi}}-\delta(1-x)\int dx'\int_0^1 dy {x'+y\over 1-x'}{1\over |x'-y|^{1+2\epsilon}}{\Gamma(\epsilon+{1\over2})\over\sqrt{\pi}} \right]\nonumber\\
	&= {\alpha_sC_F\over 2\pi}\left({\mu^2\over p_z^2}\right)^\epsilon\left[\int_0^1 dy {x+y\over 1-x}{1\over |x-y|^{1+2\epsilon}}{\Gamma(\epsilon+{1\over2})\over\sqrt{\pi}}+\delta(1-x)\left({1\over\epsilon_{\mbox{\tiny UV}}} - {1\over \epsilon_{\mbox{\tiny IR}}}\right)\right.\nn\\
	&\qquad\left. -\delta(1-x)\int dx'\int_0^1 dy {1+y\over 1-x'}{1\over |x'-y|^{1+2\epsilon}}{\Gamma(\epsilon_{\mbox{\tiny IR}}+{1\over2})\over\sqrt{\pi}}\right]\,,
	\end{align}
and for the tadpole diagram
	\begin{align}
	&\tilde{q}^{(1)}_\text{tadpole}(x,p^z,\epsilon)\nn\\
	=& {\alpha_sC_F\over 2\pi}\left({4\pi\mu^2\over p_z^2}\right)^\epsilon \left[-{1\over 1-2\epsilon}{1\over |1-x|^{1+2\epsilon}}{\Gamma(\epsilon+{1\over2})\over\sqrt{\pi}} -\delta(1-x) {1\over 1-2\epsilon}{1\over |1-x|^{1+2\epsilon}}\int_{-\infty}^\infty dx' {1\over |1-x'|^{1+2\epsilon}}{\Gamma(\epsilon+{1\over2})\over\sqrt{\pi}}\right]\nonumber\\
	=& {\alpha_sC_F\over 2\pi}\left({4\pi\mu^2\over p_z^2}\right)^\epsilon \left[-{1\over 1-2\epsilon}{1\over |1-x|^{1+2\epsilon}}{\Gamma(\epsilon+{1\over2})\over\sqrt{\pi}} +\delta(1-x)\left({1\over\epsilon_{\mbox{\tiny UV}}} - {1\over \epsilon_{\mbox{\tiny IR}}}\right) \right]\,.
	\end{align}
	
\end{widetext}
After adding these expressions and carrying out the remaining integrations over $y$, we obtain the same result as in Eq.~(\ref{eq:1loopquasi}).

\section{$\epsilon$ Expansion and Plus Functions}
\label{sec:ep}

Since the support of the quasi-PDF ranges from $-\infty$ to $\infty$, its asymptotic behavior as $\sim1/|x|$ at $|x|\to\infty$ implies a UV divergence which can be regularized by dimensional regularization. Therefore, the $\epsilon$ expansion of the quasi-PDF should account for this feature.

In general, we need to expand
\beq
	{\theta(x)\over x^{1+\epsilon}} = {\theta(x)\theta(1-x)\over x^{1+\epsilon}} + {\theta(x-1)\over x^{1+\epsilon}}\,,
\eeq
and it is well known that
\begin{align} \label{eq:ident}
{\theta(x)\theta(1-x)\over x^{1+\epsilon}} = - {1\over\epsilon}\delta(x)+
\tilde{\mathcal{L}}_0(x) - \epsilon \tilde{\mathcal{L}}_1(x)+ O(\epsilon^2)\,,
\end{align}
where $\epsilon<0$.
Here we follow~\cite{Ligeti:2008ac} and the plus functions $\mathcal{L}_n(x)$ are defined as
\begin{align} \label{eq:plus1}
	\mathcal{L}_n(x)\equiv &\left[ {\theta(x)\ln^nx\over x}\right]_+
   \\
	= &\lim_{\beta\to0} \left[{\theta(x-\beta)\ln^n x\over x} + \delta(x-\beta) {\ln^{n+1}\beta \over n+1}\right]
 \,, \nn
\end{align}
and we let
\begin{align}
  \tilde{\mathcal{L}}_n(x) = \theta(1-x) \mathcal{L}_n(x) \,.
\end{align}
Note that  $\int_0^1 dx \, {\cal L}_n(x) =0$.

To define the expansion in the range $x\in [1,\infty]$ we simply map this interval to $[0,1]$ via $t=1/x$. Taking an arbitrary smooth test function $g(x)$ we have
\begin{align}
	&\int_1^\infty dx {1\over x^{1+\epsilon}} \: g(x)
    \\
	=& \int_0^1 dt {1\over t^{1-\epsilon}}  \:  g(1/t)
   \nn\\
	=& \int_0^1 dt \left[{1\over\epsilon}\delta(t)+ \mathcal{L}_0(t) + \epsilon \mathcal{L}_1(t)+ O(\epsilon^2)\right] \, g(1/t)
    \nn\\
	=& \int_1^\infty\!\! {dx\over x^2}\left[{1\over\epsilon}\delta^+\Bigl({1\over x}\Bigr)+ \mathcal{L}_0\Bigl({1\over x}\Bigr) + \epsilon \mathcal{L}_1\Bigl({1\over x}\Bigr)+ O(\epsilon^2)\right] g(x)
   \,.\nn
\end{align}
Here $\epsilon>0$ and the superscript $+$ on the $\delta^+$ function indicates that its argument should be positive. Therefore $\delta^+(1/x)$ has its support at $x=+\infty$, not $x=-\infty$.  Since $g$ is arbitrary we can identify
\begin{align}
  {\theta(x-1)\over x^{1+\epsilon}}
   &= {1\over\epsilon}{1\over x^2}\delta^+\Bigl({1\over x}\Bigr)
   + \frac{1}{x^2} \tilde{\mathcal{L}}_0\Bigl({1\over x}\Bigr)
   + \epsilon \frac{1}{x^2} \tilde{\mathcal{L}}_1\Bigl({1\over x}\Bigr)
  \nn\\
  &\quad + O(\epsilon^2)\,.
\end{align}
Combining the above results and denoting which $1/\epsilon$ poles are UV or IR divergences we have
\begin{align} \label{eq:expansion}
	{\theta(x)\over x^{1+\epsilon}}
 = & \left[-\frac{1}{\epsilon_{\mbox{\tiny IR}}}\delta(x) + \frac{1}{\epsilon_{\mbox{\tiny UV}}} {1\over  x^2}\delta^+\Bigl({1\over x}\Bigr)\right]
    \\
    & + \left({1\over x}\right)^{[0,1]}_{+(0)}+\left({1\over x}\right)^{[1,\infty]}_{+(\infty)}
  \nn\\
	& - \epsilon\left[\left({\ln x\over x}\right)^{[0,1]}_{+(0)}+\left({\ln x\over x}\right)^{[1,\infty]}_{+(\infty)}\right] + O(\epsilon^2)
   \,, \nn
\end{align}
where we have defined the distributions
\begin{align}
  \big(1/x\big)_{+(0)}^{[0,1]} &\equiv  \tilde{{\cal L}}_0(x)
  \,,  \quad
  \big(\ln x/x\big)_{+(0)}^{[0,1]} \equiv  \tilde{{\cal L}}_1(x)
  \,,
\end{align}
and
\begin{align}\label{eq:plusinfinity}
 \big(1/x\big)_{+(\infty)}^{[1,\infty]} &\equiv (1/x^2) \tilde{{\cal L}}_0(1/x)
  \,,\nn \\
\big(\ln x/x\big)_{+(\infty)}^{[1,\infty]} &\equiv -(1/x^2) \tilde{{\cal L}}_1(1/x)
  \,.
\end{align}
Note that \eq{expansion} is consistent with the expected result that
\begin{align}
 \int_0^\infty \frac{dx}{x^{1+\epsilon}}
 = \frac{1}{\epsilon_{\mbox{\tiny UV}} }
  -\frac{1}{\epsilon_{\mbox{\tiny IR}} } \,.
\end{align}

\bibliography{ref}

\end{document}